\documentclass[sigconf,screen]{acmart}

\AtBeginDocument{%
  }

\usepackage{xspace}
\usepackage{url}

\usepackage{microtype}
\usepackage[framemethod=tikz]{mdframed}
\newcounter{finding}
\newcommand{\finding}[1]{\refstepcounter{finding}
  \vspace{2.3mm}
 \begin{mdframed}[linecolor=gray,roundcorner=12pt,backgroundcolor=gray!15,linewidth=3pt,innerleftmargin=2pt, leftmargin=0cm,rightmargin=0cm,topline=false,bottomline=false,rightline = false]
  \textbf{Finding \arabic{finding}:} #1
 \end{mdframed}
 \vspace{2.3mm}
}

\usepackage{cancel}
\usepackage{enumitem}

\usepackage{graphicx}
\usepackage{subfigure}

\setcopyright{cc}
\setcctype{by}
\acmDOI{10.1145/3832783.3834343}
\acmYear{2026}
\copyrightyear{2026}
\acmISBN{979-8-4007-2882-2/2026/10}
\acmConference[ASE '26]{Proceedings of the 41st IEEE/ACM International Conference on Automated Software Engineering}{October 12--16, 2026}{Munich, Germany}
\acmBooktitle{Proceedings of the 41st IEEE/ACM International Conference on Automated Software Engineering (ASE '26), October 12--16, 2026, Munich, Germany}
\acmSubmissionID{ase26main-p161-p}
\received{2026-03-26}
\received[accepted]{2026-06-18}

\begin{document}

\title{Characterizing the Landscape of Open-Source Satellite Software}

\author{Jinfeng Wen}
\orcid{0000-0003-3023-1005}
\affiliation{%
  \institution{Beijing University of Posts and Telecommunications}
  \city{Beijing}
  \country{China}
}
\email{jinfeng.wen@bupt.edu.cn}

\author{Qi Liang}
\orcid{0009-0001-1249-8905}
\affiliation{%
  \institution{Beijing University of Posts and Telecommunications}
  \city{Beijing}
  \country{China}
}
\email{qliang@bupt.edu.cn}

\author{Yuehan Sun}
\orcid{0009-0009-5803-6907}
\affiliation{%
  \institution{Beijing University of Posts and Telecommunications}
  \city{Beijing}
  \country{China}
}
\email{sunyuehangm24@163.com}

\author{Federica Sarro}
\orcid{0000-0002-9146-442X}
\affiliation{%
  \institution{University College London}
  \city{London}
  \country{United Kingdom}
}
\email{f.sarro@ucl.ac.uk}

\author{Ao Zhou}
\orcid{0000-0001-5743-9418}
\affiliation{%
  \institution{Beijing University of Posts and Telecommunications}
  \city{Beijing}
  \country{China}
}
\email{aozhou@bupt.edu.cn}

\author{Xuanzhe Liu}
\orcid{0000-0002-7908-8484}
\affiliation{%
  \institution{Peking University}
  \city{Beijing}
  \country{China}
}
\email{liuxuanzhe@pku.edu.cn}

\author{Shangguang Wang}
% \authornote{Corresponding author.}
\correspondingauthor
\orcid{0000-0001-7245-1298}
\affiliation{%
  \institution{Beijing University of Posts and Telecommunications}
  \city{Beijing}
  \country{China}
}
\email{sgwang@bupt.edu.cn}

\renewcommand{\shortauthors}{Jinfeng Wen et al.}

\begin{abstract}
Satellites have become fundamental components of modern technological systems, supporting critical infrastructure in communication, navigation, Earth observation, and scientific research. As space exploration advances and demand for satellite-enabled services grows, reliance on complex, heterogeneous satellite software continues to increase. A systematic understanding of the satellite software landscape is therefore increasingly important, yet existing studies still lack a comprehensive empirical examination. To address this gap, we present the first characterization study of open-source satellite software, examining its ecosystem and development practices. We mine and analyze 22,286 satellite-related GitHub projects through three research questions on popularity trends (RQ1), software goals (RQ2), and development practices (RQ3). First, we characterize the temporal evolution of projects and active developers, revealing increasing popularity. Second, through manual inspection of 646 projects, we construct a taxonomy of 43 software-goal categories spanning conceptual design, datasets, system implementation, simulation, testing, and tools. Third, we conduct an in-depth analysis of projects with source code, revealing a highly heterogeneous and task-specialized ecosystem with 66 programming languages and diverse implementation strategies. Finally, we summarize key findings and derive actionable implications for satellite developers and researchers.

\end{abstract}





\maketitle

\section{Introduction}\label{sec:introduction}

% Satellite popular and critical

% With the rapid advancement of aerospace technology, satellites have become a fundamental infrastructure of modern society. Satellite technology has made significant improvements across commercial, civil, and military fields. Large-scale satellite constellations, consisting of hundreds to tens of thousands of satellites, such as SpaceX Starlink~\cite{}, OneWeb~\cite{}, Amazon Kuiper, and Telesat~\cite{}, are being actively deployed worldwide. They provide essential support for a wide range of applications, including high-resolution Earth observation~\cite{}, precise navigation~\cite{}, low-latency communications~\cite{}, and scientific research~\cite{}. 

With the rapid advancement of aerospace and information technologies, satellites have transitioned from isolated technical assets into indispensable infrastructure supporting modern society. The advent of large-scale satellite constellations has fundamentally transformed the architecture and operational paradigm of space systems. Commercial mega-constellations, e.g., SpaceX Starlink~\cite{IntroStarlink}, OneWeb~\cite{IntroOneWeb}, and Amazon Kuiper~\cite{IntroKuiper}, envision the deployment of hundreds to tens of thousands of satellites in Low Earth Orbit (LEO). As a representative case, Starlink has already launched over 9,000 satellites, with longer-term plans to expand the constellation to approximately 42,000 satellites~\cite{starlinkdetail}. Beyond LEO, contemporary satellite infrastructures also encompass Medium Earth Orbit (MEO) and Geostationary Earth Orbit (GEO) systems, offering complementary advantages in coverage, capacity, and service continuity.
At present, satellites underpin a broad spectrum of applications, including high-resolution Earth observation~\cite{earthobservation, zhai2023fedleo}, global navigation and positioning services~\cite{khalife2021first}, low-latency broadband communications~\cite{ma2023network, mohan2024multifaceted}, and scientific exploration missions~\cite{wang2021tiansuan, wen2025satelight}.

Given the growing prominence of the satellite domain, related studies have primarily concentrated on satellite network design and optimization~\cite{pan2023pmsat, singh2021community, feng2024distributed}, network attack detection and mitigation~\cite{giuliari2021icarus, usman2020mitigating}, task and resource scheduling~\cite{zhai2024seco, liu2024orbit, xie2024computation}, satellite attitude control and adjustment~\cite{robic2022vision, legrand2022end}, in-orbit task processing~\cite{denby2023kodan, zhang2024resource}, satellite imagery analysis~\cite{hu2020image, de2021rainbench}, orbital system design~\cite{denby2020orbital}, etc. In addition, several work has investigated the measurement and empirical analysis of satellite-related systems and data, including real-device measurements of satellite systems~\cite{xing2024deciphering}, analyses of communication transmission performance~\cite{pan2022latency}, investigations of satellite modem security~\cite{yu2024comprehensive}, measurement of Starlink network performance~\cite{ma2023network, michel2022first, mohan2024multifaceted, izhikevich2024democratizing}, label analyses of onboard data~\cite{yang2025fedclr}, etc. 
Beyond these efforts, software engineering (SE) researchers have increasingly turned their attention to the satellite domain, addressing specific tasks, e.g., onboard software update~\cite{wen2025satelight}, equipment code generation~\cite{he2026enhancing}, system design and configuration~\cite{gios2024vision, flederer2021configurable, apvrille2004verifying}, testing scheduling and prioritization~\cite{ollando2026test, shin2018test}, formal modeling~\cite{esteve2012formal}, and operational planning~\cite{ozturk2023software}.

% In modern missions, satellite software has evolved into a highly complex and mission-critical system, responsible for communication management, attitude and orbit control, system monitoring, telecommand processing, telemetry acquisition, and ground–satellite coordination. 

Despite these advances, existing studies have largely overlooked a systematic understanding of the satellite software landscape. Analyses of historical satellite missions indicate persistent and severe reliability challenges associated with software-intensive systems. Statistics reported by NASA~\cite{failurerate} show that approximately 24.2\% of satellite missions suffered total failure, while an additional 11\% experienced partial failure. Further investigations reveal that a substantial fraction of these failures cannot be attributed solely to hardware faults~\cite{rico2016combined}. Instead, they are closely linked to software behavior, design defects, or unexpected interactions. Satellite software is responsible for core mission execution logic, attitude and orbit control, telecommand processing, and autonomous decision-making~\cite{diversesoftware}. Given its central role in coordinating mission-critical functions, satellite software has become a primary determinant of overall mission reliability and operational lifetime.

With the advancement of space exploration and the commercialization of aerospace systems, satellite software development has become increasingly open and collaborative~\cite{satelliterepo}. Open implementations now exist for core functions such as attitude control~~\cite{p301}, orbit propagation~\cite{p118}, communication signal control~\cite{p20}, constellation simulation~\cite{p310}, and satellite data processing~\cite{p285}. This trend highlights the growing role of open-source software (OSS) in satellite engineering. Studying these OSS-based satellite projects is therefore valuable for academia and practice, as it enables knowledge sharing, code reuse, lower development costs, and faster technological progress. However, to the best of our knowledge, no prior work has systematically examined OSS satellite projects.

To bridge this gap, we present the first comprehensive study of OSS satellite projects. We analyze OSS satellite projects hosted on GitHub, the world's largest and most influential open-source platform~\cite{businge2019studying, bao2019large, pickerill2020phantom}. Owing to its diversity and representativeness, GitHub provides a reliable basis for empirical software engineering research~\cite{aghili2023studying, ramasamy2023workflow, das2022empirical, trockman2019striking, businge2019studying, gonzalez2020state}. We collect and examine 22,286 OSS satellite projects to address three research questions.

$\bullet$ \textbf{RQ1 (Popularity Trend):} We examine the popularity trend of OSS satellite projects. The results reveal a sustained and significant growth in community attention to satellite software, indicating the increasing prominence of satellite-related topics and underscoring the timeliness and necessity of this study.

% The results show a significant increase in community attention toward satellite software. This trend confirms the growing importance of satellite topics and highlights the timeliness and urgency of our study.

$\bullet$ \textbf{RQ2 (Goal Taxonomy):} We investigate the functional objectives that satellite projects are designed to achieve. 
% We randomly sample 646 relevant projects for manual inspection. For each project, we identify its underlying development goal.
A random sample of 646 projects is manually inspected to identify development goals.
Based on this analysis, we construct a taxonomy consisting of 43 goal categories. The resulting taxonomy shows that OSS satellite projects span a broad range of objectives, including datasets, system implementation, simulation, testing, and diverse tools.

% linked to goals in developing satellite projects. It indicates that satellite software has a wide spectrum of objectives in satellites, covering data acquisition, system implementation, simulation, testing, and auxiliary tools, etc.

$\bullet$ \textbf{RQ3 (Development Practices):} We examine development practices through a systematic analysis of project source code. The results identify the use of 66 distinct programming languages, with language selection closely aligned with specific task requirements. In addition, implementation strategies increasingly emphasize the integration of heterogeneous models and solutions.

% The results show the adoption of 66 distinct programming languages, with language choices closely aligned with task requirements. Moreover, implementation strategies are increasingly covering the integration of diverse models or solutions.

% Our analysis reveals the use of 66 distinct programming languages, selected in a task-aligned manner. Implementation decisions increasingly converge on the integration of diverse model choices.

Based on these analyses, we derive key findings and actionable implications for satellite software developers and researchers.

% In addition, we publicly release the scripts and dataset in this study~\cite{ourdata}, thereby enabling replication and supporting future research.

% In addition, we offer the scripts and the dataset used in our study~\cite{ourdata} as an n, we offer the scripts and the dataset used in this study as an additional contribution to the research community for other researchers to replicate and build upon.

% The results indicate a clear trend toward heterogeneous and task-specialized development ecosystems. Language and implementation decisions increasingly converge on pragmatic task alignment and the integration of machine-learning–driven enhancement. 

% A clear shift toward the implementation paradigm is shown, coupling physics-based modeling with machine-learning–driven enhancement.

\section{Background and Related Work}\label{sec:relatedwork}

\noindent \textbf{Satellite Software vs. Traditional Software.} Unlike traditional terrestrial software, satellite software is developed under distinctive environmental, operational, and mission constraints. These constraints shape its design, implementation, and execution.
First, satellite software operates in highly resource-constrained environments. Limited power, thermal capacity, and onboard hardware impose strict bounds on computation and storage. As a result, its design priorities may differ from those of traditional software. Some software requires direct support of critical functions, e.g., flight control and task management, while others closely interact with onboard hardware components.
Second, satellite software relies on specialized communication mechanisms rather than the unified TCP/IP-based infrastructure common in terrestrial software. Developers may consider more low-level communication issues and accommodate unstable links and long delays.
Last but not least, the satellite software ecosystem may be less standardized and less mature than that of traditional software. 
% Toolchains are fragmented and mission-specific. 
% Testing support is limited.
% , which slows development and complicates validation. 
Moreover, because space missions are costly and high-risk, satellite software may depend on high-fidelity simulation before deployment.

These characteristics distinguish satellite software from traditional software, yet its overall landscape remains insufficiently understood. A systematic characterization is therefore needed to clarify its current state and provide actionable insights.

\noindent \textbf{Related Studies on Satellites.} Existing satellite studies have conducted extensive empirical analyses across multiple dimensions, including Starlink network characteristics and performance through large-scale measurements~\cite{ma2023network, michel2022first, mohan2024multifaceted, izhikevich2024democratizing}, real-device measurements of satellite platforms~\cite{xing2024deciphering}, communication latency and reliability~\cite{pan2022latency}, security vulnerabilities in satellite communication modems~\cite{yu2024comprehensive}, and the effects of onboard data constraints on learning performance~\cite{yang2025fedclr}.
% Existing studies on satellites have conducted extensive empirical analyses across multiple dimensions. They have investigated Starlink network characteristics and performance through large-scale measurements~\cite{ma2023network, michel2022first, mohan2024multifaceted, izhikevich2024democratizing}, real-device measurements of satellite platforms~\cite{xing2024deciphering}, communication transmission latency and reliability~\cite{pan2022latency}, security vulnerabilities in satellite communication modems~\cite{yu2024comprehensive}, onboard data constraints on learning performance~\cite{yang2025fedclr}, and so on. 
Specifically, Ma~\textit{et al.}~\cite{ma2023network} analyzed end-to-end network characteristics of Starlink from an end-user perspective. 
% Mohan~\textit{et al.}~\cite{mohan2024multifaceted} presented a comprehensive multi-faceted performance analysis of Starlink using diverse measurement sources. 
Xing~\textit{et al.}~\cite{xing2024deciphering} examined how thermal control and power management affect computing capabilities and task scheduling on commercial off-the-shelf satellite devices. 
% Pan~\textit{et al.}~\cite{pan2022latency} characterized end-to-end delays in inter-LEO and LEO–terrestrial communication architectures. 
Yu~\textit{et al.}~\cite{yu2024comprehensive} performed a study of satellite modems and revealed potential security risks. 
Yang~\textit{et al.}~\cite{yang2025fedclr} analyzed satellite-ground federated learning and showed that data label scarcity and skewness significantly degrade model accuracy.
% with more severe conditions leading to larger performance losses.
% Overall, these studies primarily focus on characteristics of satellite networks, platform devices, communication performance, security, and data labeling. In contrast, satellite software, especially its open-source ecosystem and characteristics, has received limited attention. Our work addresses this gap.
Overall, prior studies focus on satellite networks, devices, communication, security, and data labeling, whereas open-source satellite software remains underexplored. Our work fills this gap.

\noindent \textbf{Related Studies on Open-Source Software.}
A large body of research has examined the characteristics of OSS, with GitHub repositories serving as a primary data source. These studies have spanned diverse domains, including AIOps~\cite{aghili2023studying}, data science~\cite{ramasamy2023workflow}, Blockchain~\cite{das2022empirical, trockman2019striking}, developer social networks~\cite{wang2022quantifying, tamburri2019exploring}, machine learning~\cite{gonzalez2020state}, mobile app economy~\cite{businge2019studying}, etc.
For instance, Aghili~\textit{et al.}~~\cite{aghili2023studying} studied OSS AIOps projects, focusing on data types and analysis techniques. Ramasamy~\textit{et al.}~\cite{ramasamy2023workflow} investigated data science practices, examining common activities, workflow stages, and transitions between stages. Das~\textit{et al.}~\cite{das2022empirical} empirically studied the Blockchain projects, with an emphasis on specific project categories and user engagement patterns. Businge~\textit{et al.}~\cite{businge2019studying} analyzed the factors influencing the popularity of mobile applications.
% Despite extensive research across multiple domains, open-source software in the satellite domain remains largely unexplored. 
% Satellites are increasingly pervasive and play a critical role in modern society. Thus, we present the first comprehensive empirical study of open-source satellite projects, demystifying their popularity trends, the landscape of software goals, and corresponding development practices, as well as providing promising insights.
% Despite extensive research across multiple domains, OSS satellite projects remain largely unexplored. As satellites become increasingly pervasive and critical to modern society, a systematic understanding of their software ecosystem is urgently needed. We present the first comprehensive empirical study of OSS satellite projects, providing a holistic characterization of this emerging domain and insights for future research and practice.
% Despite extensive research in related domains, OSS satellite projects remain largely unexplored. As satellites become increasingly critical to modern society, an understanding of their software ecosystem is imperative. We present the first comprehensive study of OSS satellite projects, offering a holistic characterization and actionable implications for future research and practice.
Despite extensive research in related domains, OSS satellite projects remain unexplored. As satellites become increasingly critical to modern society, understanding their software ecosystem is imperative. We present the first characterization study of OSS satellite projects.

% Additionally, some studies have focused on the specific properties of open-source projects, including development effort estimation~\cite{robles2022development, jorgensen2006systematic}, XX, and XX. For instance, 

% Robles~\textit{et al.}~\cite{robles2022development} analyzed historical data of developers' effort to develop a simple effort estimation model for OSS projects.

\section{Methodology}\label{sec:methodology}

% To comprehensively understand the characteristics of OSS satellite projects, this study presents an overview of the methodology, as shown in Fig.~\ref{fig:methodology}. We first present and aim to answer the following three research questions.

To characterize OSS satellite projects, we follow previous studies~\cite{bock2023automatic, chen2020comprehensive, lou2020understanding, humbatova2020taxonomy, garcia2012survey, WenServerless21} to guide data collection and analysis, and to define the research questions addressed in this study.

% To characterize OSS satellite projects, we collect the data and 

% we provide an overview of the research methodology, as illustrated in Fig.~\ref{fig:methodology}. We first present and seek to answer the following three research questions.

% \begin{figure}[t]
% 	\centering
%     \includegraphics[width=0.98\textwidth]{pic/methodology.pdf}
%     % \vspace{-3mm}
%     \caption{An overview of our methodology.}
%     % \vspace{-1mm}
%     \label{fig:methodology}
%     \vspace{-3mm}
% \end{figure}

% \subsection{Research Questions}

% \textbf{\textit{RQ1: Determine Popularity Trend (When)}}: What is the trend of popularity of open-source satellite software?

% \textbf{\textit{RQ2: Determine Taxonomy of Software Goal (What)}}: What are the specific goals that developers aim to achieve in open-source satellite software?

% \textbf{\textit{RQ3: Determine Current Practice (How)}}: What methods and analytical techniques, as well as programming languages, are predominantly employed in satellite software that provides source code?

$\bullet$ \textbf{RQ1: Popularity Trend.} \textit{What is the popularity trend of OSS satellite projects?}

% \textbf{RQ1: Popularity Trend.} \textit{How is the popularity trend of OSS satellite projects?}

$\bullet$ \textbf{RQ2: Goal Taxonomy.} \textit{What specific goals do OSS satellite projects address?}

% \textbf{\textit{RQ3 (How): Current Practice}}: What development practices, analytical methods, or programming languages are predominantly adopted in OSS satellite projects?

$\bullet$ \textbf{RQ3: Development Practices.} \textit{How are development practices adopted in OSS satellite projects?}

\subsection{Data Preparation}
% GitHub REST API

We collect OSS satellite projects from GitHub on October 10, 2025. Using the GitHub REST API~\cite{githubapi}, we query repositories with the keyword ``satellite'' and initially retrieve 32,347 projects. To ensure data quality and research relevance, we then apply an automatic filtering process to query the retrieved projects. Specifically, we exclude projects that (1) return a 404 error, indicating inaccessible or deleted projects; (2) lack essential project artifacts, i.e., README files and implementation details, which are required to infer software intent; and (3) are written in non-English languages. 
% These criteria ensure that the remaining projects contain sufficient information for empirical analysis. Finally, 22,287 OSS satellite projects (referred to as satellite software) are retained for subsequent analysis.
These criteria ensure that the retained projects provide sufficient information for empirical analysis. Ultimately, 22,286 OSS satellite projects (referred to as satellite software) are selected for subsequent analysis.

\subsection{Methodology Designs}

\noindent \textbf{Design of RQ1}. To examine the popularity trends, we follow prior work~\cite{WenServerless21, bagherzadeh2019going, chen2020comprehensive, calefato2022will, bock2023automatic} and measure the annual number of newly created projects and active developers across our dataset of 22,286 repositories. Newly created projects are identified using the \texttt{created\_at} timestamp, while annual active developers are determined from commit records by counting unique developers per year across all projects. In addition, we compute the cumulative number of active developers up to each year via developer identifier deduplication, and calculate the annual number of commits as a complementary indicator of satellite software popularity.

\noindent \textbf{Design of RQ2}. To construct a taxonomy of software goals, we analyze and categorize the functional objectives of OSS satellite projects. Given the large dataset of 22,286 projects, exhaustively labeling all projects is impractical. Following established empirical practices~\cite{icseAghajaniNVLMBL19, WenServerless21}, we randomly sample a statistically representative subset that ensures a 99\% confidence level with a margin of error of $\pm$5\%. Note that the 99\% confidence level refers to the statistical confidence for estimating the population proportion of projects in each labeled category.
This procedure yields 646 projects for taxonomy construction. Particularly, the resulting dataset is larger than those used in prior empirical studies~\cite{CHENDLDEPLOY2, isstaZhangCCXZ18, WenServerless21}, providing a stronger empirical foundation for taxonomy construction.  
We then construct the taxonomy using the open coding procedure~\cite{seaman1999qualitative}, a standard qualitative analysis technique widely adopted in empirical SE research~\cite{chen2020comprehensive, lou2020understanding, humbatova2020taxonomy, garcia2012survey, WenServerless21}, to ensure rigor and reliability.

First, we randomly sample 70\% of satellite projects (453) and inductively derive categories and subcategories in a bottom-up manner. Two authors collaboratively analyze project descriptions, README files, source code, and related files through iterative reviews to identify project goals. The procedure is as follows. Projects unrelated to satellites are labeled as \textit{False} and excluded from the taxonomy. For example, Satellite~\cite{p36} is a Red Hat enterprise IT infrastructure management tool unrelated to spaceborne satellites or satellite technologies despite its name.
% Projects unrelated to satellites are labeled as \textit{False} and excluded from the taxonomy. For instance, Red Hat Satellite~\cite{p34} is an enterprise-grade IT infrastructure management tool developed by Red Hat. Despite the inclusion of the term ``satellite'' in the project~\cite{p34}, this tool is not related to actual satellites or associated technologies. 
% Software containing only a README file without substantive content is labeled as \textit{Null}. For example, project~\cite{p315} included only a README description. 
For the remaining projects, the authors generate concise phrases summarizing the project goals. These phrases are iteratively grouped into categories, forming a hierarchical taxonomy. When a project addresses multiple objectives, it is assigned to all applicable categories. For example, one project~\cite{p118} supports both satellite coverage analysis and orbit propagation. The former evaluates access to ground locations along the orbit, whereas the latter computes position and velocity over time from initial orbital elements. Accordingly, the project is assigned to \textit{[F.2.1] Satellite Coverage Analysis} and \textit{[F.3.2] Orbit Propagation}.
% When a project aligns with multiple categories, it is assigned to all relevant categories. For example, one project~\cite{p118} implemented both satellite coverage analysis and orbit propagation tools. The satellite coverage analysis tool calculates the satellite's ability to access or observe ground locations during its orbit. The orbit propagation tool computes the position and velocity of a satellite over a specified time period based on its initial orbital parameters. Accordingly, this project is classified under two categories: \textit{Satellite Coverage Analysis (F.2.1)} and \textit{Orbit Propagation (F.3.2)}.
Any disagreements are resolved through discussion with a third arbitrator, an expert with ten years of satellite research experience. This iterative and consensus-driven process yields the initial taxonomy, with project goals validated by the authors and the arbitrator.

Second, we refine and extend the taxonomy using the remaining 30\% of projects (193). Two authors independently label these projects based on the initial taxonomy, assigning them to categories. Projects that do not fit existing categories are temporarily assigned to a \textit{Pending} category. We calculate inter-rater agreement using Cohen's Kappa ($\kappa$)~\cite{cohen1960coefficient}, achieving a value of 0.857, indicating near-perfect agreement and supporting the reliability of our labeling process~\cite{landis1977measurement}. The authors, together with an arbitrator, resolve the remaining disagreements and reclassify \textit{Pending} projects into existing or newly created categories. This process finalizes the taxonomy with the introduction of four additional categories and indicates a relative stabilization of the satellite goal taxonomy.
% The authors and arbitrator jointly resolve remaining disagreements and reclassify \textit{Pending} projects into appropriate existing or newly created categories, thereby finalizing the taxonomy with the addition of four new categories. This also indicates a relative saturation for all satellite categories.

Among 646 sampled OSS satellite projects, 34 are labeled \textit{False}, 604 are assigned to one category, and 8 to two categories. The final taxonomy comprises 620 samples.
We further analyze the popularity of the 620 samples using the same RQ1 metrics as for the full dataset. The results show similarly increasing trends in annual project creation and active developers (later reported in Section~\ref{sec:rq1}), suggesting that the observed growth is robust to sampling and not mainly driven by false positives in the full dataset.

\noindent \textbf{Design of RQ3}. 
% To investigate the current development practices associated with each satellite goal category, we conduct an in-depth analysis of the project source code. This analysis focuses on two aspects: programming language practices and implementation practices. 
To examine development practices across satellite goal categories, we conduct an in-depth analysis of source-code-available projects among the 620 samples. The analysis focuses on programming language practices and implementation practices.
For language practices, we derive language usage from the per-project language statistics, which report the proportion of each language. As many projects use multiple languages, these proportions are normalized and aggregated. For each goal category, we sum the language percentages across all projects and compute the relative share of each language within the category.
% For language practices, their usage is determined from the language statistics reported in each project, where the proportion of each programming language is specified. Since some projects show multiple programming language usage, these percentages require further normalization and aggregation. For each category with multiple projects sharing the same goal, we compute the distribution of each language by summing the language percentages across all projects in the category and calculating the share of each specific language relative to this total. 
% For implementation practices, we perform code inspection, extract relevant code segments, and review cross-reference documentation to ensure accurate interpretation of technical approaches or analytical perspectives. 
For implementation practices, we inspect source code, extract relevant segments, and review cross-referenced documentation to ensure accurate interpretation of technical approaches and analytical perspectives.

\section{RQ1: Popularity Trend}\label{sec:rq1}

% \begin{figure}[t]
%     \centering
%     % \vspace{-3mm}
%     \begin{minipage}[t]{0.45\textwidth}
%         \centering
%         \includegraphics[width=\textwidth]{pic/popularity1.pdf}
%         \caption{(RQ1) The popularity trend of annual new projects and active developers.}
%         \label{fig:popularity1}
%     \end{minipage}
%     \hfill
%    \begin{minipage}[t]{0.45\textwidth}
%         \centering
%         \includegraphics[width=\textwidth]{pic/popularity2.pdf}
%         \caption{(RQ1) The popularity trend of cumulative developers and annual commits.}
%         \label{fig:popularity2}
%     \end{minipage}
%     \vspace{-3mm}
% \end{figure}

\begin{figure}[t]
	\centering
    \includegraphics[width=0.45\textwidth]{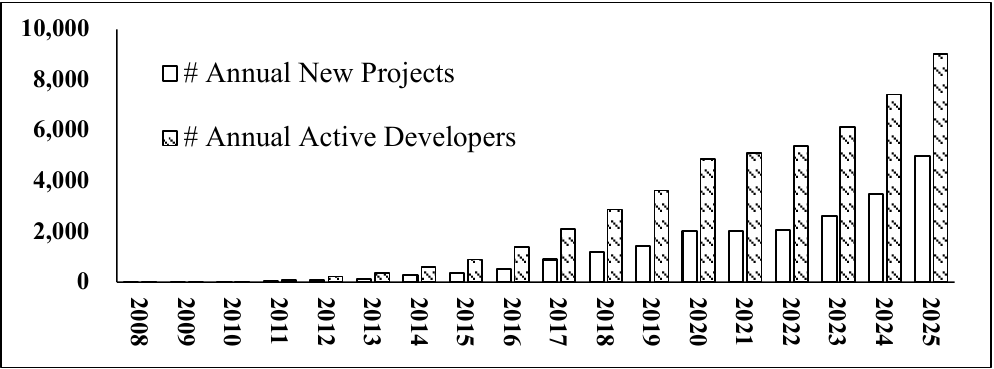}
    \vspace{-3mm}
    \caption{(RQ1) The popularity trend of annual new projects and active developers.}
    % \vspace{-1mm}
    \label{fig:popularity1}
    \vspace{-3mm}
\end{figure}

\begin{figure}[t]
	\centering
    \includegraphics[width=0.45\textwidth]{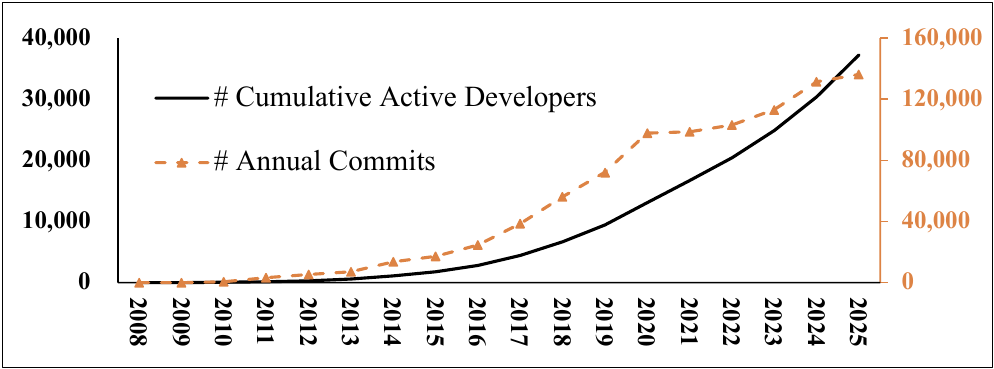}
    \vspace{-3mm}
    \caption{(RQ1) The popularity trend of cumulative developers and annual commits.}
    % \vspace{-1mm}
    \label{fig:popularity2}
    \vspace{-3mm}
\end{figure}

% Fig.~\ref{fig:popularity} shows the annual variation in the number of newly created OSS satellite projects. The results indicate a steady and pronounced upward trend from 2008 to 2025, reflecting the growing enthusiasm and sustained engagement in satellite-related topics.

% Figure 3 represents the popularity trend of serverless computing
% in terms of the number of questions and users on SO. We find
% that this topic has been gaining increasing attention since 2015,
% demonstrating the timeliness and urgency of this study.

% Fig.~\ref{fig:popularity} represents the popularity trend of satellite software through the annual number of newly created projects. Results show a steady and pronounced increase from 2008 to 2025, reflecting sustained growth in community engagement and underscoring the timeliness and relevance of this study.

Figure~\ref{fig:popularity1} illustrates the popularity trends of OSS satellite projects in terms of the annual number of newly created projects and active developers. We find that satellite software has been gaining increasing attention, underscoring the timeliness and relevance of this study.
Specifically, the number of newly created satellite projects increases from 1 in 2008 to 4,997 in 2025, representing a 42.24\% increase compared to 2024. Similarly, the number of annual active developers grows from 1 to 9,018 over the same period, reflecting the steady expansion of the contributor base; for example, 4,878 active developers are recorded in 2020, a 34.34\% increase over 2019.

Figure~\ref{fig:popularity2} further presents the cumulative number of active developers (left y-axis) and the annual number of commits (right y-axis). The cumulative developer count shows sustained growth, reaching 37,246 by the end of 2025. In terms of development activity, annual commits rise markedly, with 97,893 commits submitted in 2020 (a 35.89\% increase over 2019) and a total of 136,248 commits recorded in 2025.
These results reveal increasing openness and development activity within the satellite software ecosystem.
They also suggest that satellite software is evolving from isolated project-level efforts into a more visible ecosystem with broader participation. Thus, RQ1 establishes the timeliness of this study and highlights the need for software engineering support in this domain.

\finding{Satellite software has attracted increasing attention, underscoring the timeliness and significance of this study.}

% \finding{The sustained growth of satellite software and active developers reflects increasing community attention, emphasizing the relevance and necessity of this study.}
\vspace{-3mm}

\section{RQ2: Goal Taxonomy}\label{sec:rq2}

\begin{figure*}[t]
	\centering
    \includegraphics[width=0.96\textwidth]{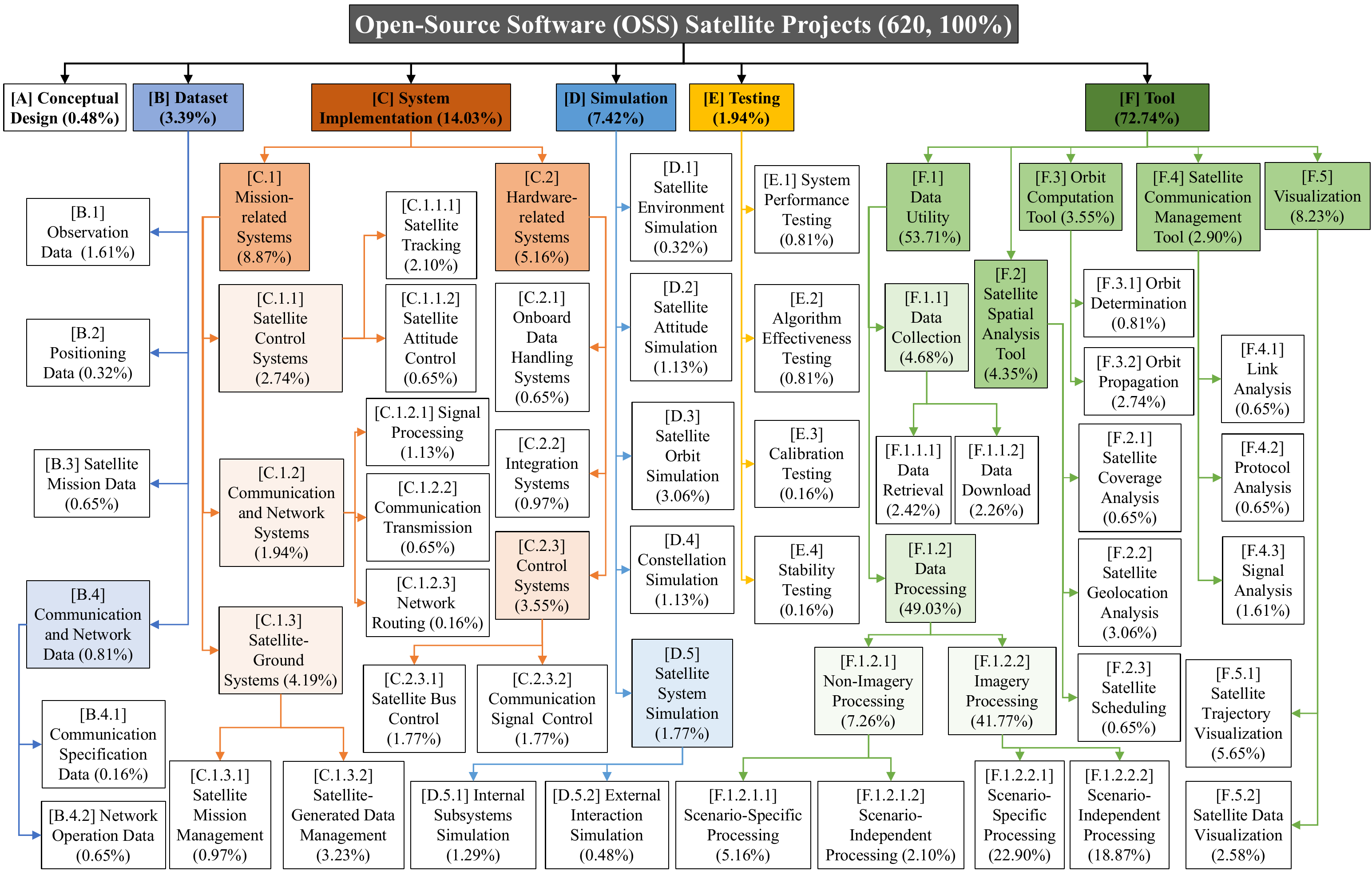}
    \vspace{-3mm}
    \caption{(RQ2) The taxonomy overview of OSS satellite project goals.}
    % \vspace{-1mm}
    \label{fig:taxonomy}
    \vspace{-3mm}
\end{figure*}

% Fig.\ref{fig:taxonomy} shows the hierarchical taxonomy of satellite projects. Nodes are descending color-level along with their depth in the hierarchy. Leaf nodes are in white. Each leaf node represents a leaf category and its non-white parent node that consists of multiple categories is an \textit{inner category}. For example, \textit{[F.4.2] Core Computation} is an inner category that can be further divided into four leaf categories: \textit{[C.4.2.1] Orbit Determination}, \textit{[C.4.2.2] Orbit Prediction}, \textit{[C.4.2.3] Orbit Propagation}, and \textit{[C.4.2.4] Orbit Perturbation Analysis}. The number of samples related to each category is in the parentheses. In total, our taxonomy consists of XX inner categories and XX leaf categories. We observe that satellite-related developers focuses on a broad spectrum of aspects, indicating a diverse attention. We next describe and exemplify each category by groups.

% Fig.~\ref{fig:taxonomy} illustrates the hierarchical taxonomy of OSS satellite projects in terms of project goals. In this taxonomy, nodes are color-coded according to their depth in the hierarchy, with colors transitioning as the levels deepen. 
Figure~\ref{fig:taxonomy} presents the hierarchical taxonomy based on satellite project goals. Nodes are color-coded by their depth in the hierarchy, with colors progressively changing across levels.
% The leaf nodes, which represent the most granular categories, are uniformly rendered in white. Each leaf node corresponds to a distinct \textit{leaf category}, while any non-white parent node encompassing multiple subcategories is referred to as an \textit{inner category}. 
Leaf nodes represent the most fine-grained categories and are uniformly rendered in white. Each leaf node corresponds to a distinct \textit{leaf category}, whereas non-white parent nodes that group multiple subcategories are defined as \textit{inner categories}.
For example, the inner category \textit{[F.3] Orbit Computation Tool} comprises two leaf categories: \textit{[F.3.1] Orbit Determination} and \textit{[F.3.2] Orbit Propagation}. 
The percentage shown in parentheses for each category denotes the proportion of satellite projects assigned to that category.
% The number in parentheses associated with each category indicates the percentage of OSS satellite projects falling under that category.
Our taxonomy comprises 22 inner categories and 43 leaf categories. The distribution demonstrates that OSS satellite projects span a wide range of objectives, reflecting a comprehensive and multi-dimensional software focus. The taxonomy captures the satellite software workflow, from conceptual design and dataset to system implementation, simulation, testing, and tool. The following sections detail each category.

\vspace{-1mm}

\finding{OSS satellite projects exhibit highly diverse objectives, spanning 22 inner categories and 43 leaf categories, and covering from \textit{Conceptual Design} and \textit{Dataset} to \textit{System Implementation}, \textit{Simulation}, \textit{Testing}, and \textit{Tool}.}
\vspace{-3mm}

\subsection{Conceptual Design (A)}

The \textit{Conceptual Design} category includes projects that introduce satellite-related concepts or designs without specifying concrete code implementations. Such projects mainly serve illustrative purposes, accounting for 0.48\% of the projects. An example is a satellite design~\cite{p420}, which is supported by multiple illustrative documents.

% The Concept Design category encompasses projects that focus on structural or model design concepts, without delving into specific implementation details. These projects are proposed by developers with an educational or illustrative intent, aiming to convey architectural ideas for satellites. For instance, one project presents the conceptual design of a space module~\cite{p277}, accompanied by comprehensive uploaded documentation to facilitate understanding of its construction. Our result shows that 3 out of 609 projects (0.49\%) fall into this category.

% focuses on the structure design concept or model concept, involving no specific implementation details, which are proposed by developers to provide educational intent. For example, the developer showed a space module concept design~\cite{project277}, along with an uploaded detailed documentation to help understand construction. Our results show that 3 projects of 610 projects (0.49\%) belong to this category

\subsection{Dataset (B)}

The \textit{Dataset} category comprises projects that provide satellite-related datasets. It accounts for 3.39\% of the surveyed projects. These datasets offer accessible resources for developers and researchers to support analysis and experimentation. They can be classified into four primary types.
% : satellite-captured observation data, positioning data derived from satellite signals, onboard-generated mission data, and communication and network data transmitted through satellite systems.
% observation data collected by satellites, positioning data derived from satellite signals, mission data generated onboard, and communication or network data transmitted through satellite systems.  
% The four types are described as follows. 
\textbf{Observation Data (B.1, 1.61\%)} refers to remote-sensing images or videos, including land-cover imagery, river inundation imagery, and infrared videos for small-target detection. 
\textbf{Positioning Data (B.2, 0.32\%)} primarily includes Global Navigation Satellite System (GNSS) datasets that support satellite navigation and geolocation analysis. 
\textbf{Satellite Mission Data (B.3, 0.65\%)} comprises onboard datasets used for satellite planning and operations, including launch history records, energy monitoring data, and mission scheduling requests.
% Examples include satellite launch history records, real-time energy monitoring data, and datasets related to mission scheduling requests and configuration management. 
\textbf{Communication and Network Data (B.4, 0.81\%)} comprises two subtypes. \textit{Communication Specification Data (B.4.1, 0.16\%)} describes satellite communication band parameters. \textit{Network Operation Data (B.4.2, 0.65\%)} includes datasets on network security and associated resource collections.
% : communication specification data and network operation data. 
% \textit{Communication Specification Data (B.4.1)} accounts for 0.16\% of the dataset describing satellite communication band parameters. \textit{Network Operation Data (B.4.2)} encompasses 0.65\% of the datasets related to network security, network resource collections, and network-simulator resources.

% \finding{A total of 3.39\% of OSS satellite projects provide dataset resources, classified into four categories. \textit{Observation Data} constitute the largest subcategory, offering remote-sensing datasets collected by satellites.
% }

\finding{3.39\% of OSS satellite projects provide dataset resources.
\textit{Observation Data} forms the largest subcategory and provides satellite-captured remote-sensing datasets.}

\vspace{-3mm}

\subsection{System Implementation (C)}

The \textit{System Implementation} category includes system software for satellite operation, control, and management. It accounts for 14.03\% of the surveyed projects and comprises two aspects. \textit{Mission-related Systems} supports high-level mission functions and abstracts from hardware. \textit{Hardware-related Systems} provides low-level support and interfaces, interacting with satellite or ground hardware.

\subsubsection{Mission-related Systems (C.1, 8.87\%)} 
\begin{figure}[t]
	\centering
    \includegraphics[width=0.45\textwidth]{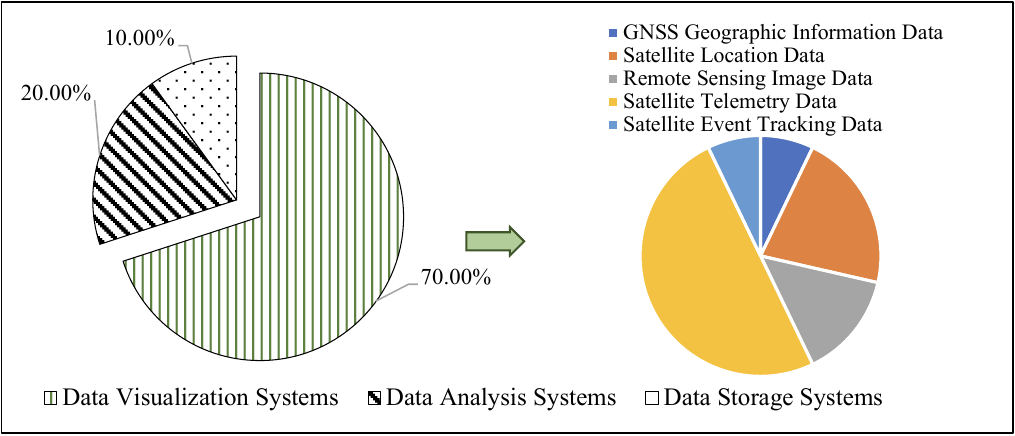}
    \vspace{-3mm}
    \caption{(RQ2) System distribution in C.1.3.2.}
    % \vspace{-1mm}
    \label{fig:C132}
    \vspace{-3mm}
\end{figure}

This category encompasses three primary types: satellite control operations, onboard communication and network management, and data connectivity between satellites and ground infrastructure. \textbf{Satellite Control Systems (C.1.1, 2.74\%)} manage satellite position and orientation in space. It comprises two subtypes: \textit{Satellite Tracking (C.1.1.1, 2.10\%)} and \textit{Satellite Attitude Control (C.1.1.2, 0.65\%)}.
Satellite tracking systems determine and monitor orbital positions. 
% Satellite tracking systems determine and monitor orbital positions through the acquisition and prediction of orbital elements. 
Satellite attitude control systems regulate and maintain satellite orientation. 
\textbf{Communication and Network Systems (C.1.2, 1.94\%)} support communication and network management across physical, link, and higher-layer protocols. It includes three types: \textit{Signal Processing (C.1.2.1, 1.13\%)}, \textit{Communication Transmission (C.1.2.2, 0.65\%)}, and \textit{Network Routing (C.1.2.3, 0.16\%)}. Signal processing systems handle signal control, reception, and decoding. Communication transmission systems improve data-transfer reliability through protocol improvement and error correction. Network routing systems address load balancing and routing strategies within satellite networks. 
\textbf{Satellite-Ground Systems (C.1.3, 4.19\%)} facilitate data exchange and operational coordination between satellites and ground systems. It represents the largest portion of category C.1 and includes two types: \textit{Satellite Mission Management (C.1.3.1, 0.97\%)} and \textit{Satellite-Generated Data Management (C.1.3.2,  3.23\%)}. 
Satellite mission management systems support mission planning, control, and monitoring. Satellite-generated data management systems support remote-sensing analysis, visualization, and storage. Figure~\ref{fig:C132} shows the distribution of C.1.3.2 systems, of which data visualization systems account for 70.00\%, mainly covering geographic information, satellite positioning, remote-sensing imagery, satellite telemetry (the dominant type), and event-tracking data.

% The system distribution in C.1.3.2 is shown in Fig.~\ref{fig:C132}. Data visualization systems constitute 70.00\% of C.1.3.2 and enable the visualization of geographic information, satellite positioning data, remote-sensing imagery, satellite telemetry data (the dominant share), and satellite event-tracking data.

\subsubsection{Hardware-related Systems (C.2, 5.16\%)}

This category involves three types.
% onboard data processing, component integration, and control subsystem optimization. 
\textbf{Onboard Data Handling Systems (C.2.1, 0.65\%)} implement onboard processing modules embedded within satellite buses using dedicated hardware boards. 
\textbf{Integration Systems (C.2.2, 0.97\%)} integrate additional hardware support into satellites. \textbf{Control Systems (C.2.3, 3.55\%)} are the largest subcategory in C.2, comprising \textit{Satellite Bus Control (C.2.3.1, 1.77\%)} and \textit{Communication Signal Control (C.2.3.2, 1.77\%)}. Based on hardware implementations, the former regulates power, thermal, and propulsion control; the latter processes communication signals via transceiver modules.

% . They control data flows through hardware-based implementations

\vspace{-1mm}

\finding{
\textit{System Implementation} accounts for 14.03\% of the surveyed projects and consists of \textit{Mission-related Systems} and \textit{Hardware-related Systems}. \textit{Mission-related Systems}, the larger subcategory (8.87\%), focuses on high-level mission functions beyond direct hardware interaction.
% 14.03\% of the projects fall under \textit{System Implementation}, which comprises \textit{Mission-related Systems} and \textit{Hardware-related Systems}. \textit{Mission-related Systems} form the largest subcategory, accounting for 8.87\% of all projects, focusing on high-level mission functions abstracted from direct hardware interaction. They primarily support data exchange and operational coordination between satellites and ground systems.
}

\vspace{-3mm}

\subsection{Simulation (D)}

The \textit{Simulation} category includes projects that model and emulate satellite-related systems, processes, or environments under controlled conditions. 
% Its primary purpose is to reduce risk and uncertainty before the deployment of algorithms or operational systems. 
This category accounts for 7.42\% of the projects, indicating the importance of experimental validation in satellite software design and evaluation. Five simulation types are employed.
% : environmental conditions, attitude dynamics, orbital motion, multi-satellite constellations, and system-level capabilities.
% Simulation serves as a key mechanism for mitigating risk and uncertainty prior to the deployment of algorithms or systems. 7.42\% of the projects fall into this category. These simulations underscore the necessity of conducting applicable experiments to inform subsequent system design and performance evaluation. The projects employ five primary simulation modalities: satellite environmental conditions, satellite attitude dynamics, orbital behavior, multi-satellite constellation configurations, and subsystem capability modeling.
\textbf{Satellite Environment Simulation (D.1, 0.32\%)} models the physical conditions affecting satellite operations, including electric fields and thermal environments. \textbf{Satellite Attitude Simulation (D.2, 1.13\%)} reproduces the dynamics and control processes of satellite attitude determination and control. \textbf{Satellite Orbit Simulation (D.3, 3.06\%)} is the largest subcategory. It generates orbital trajectories and models control procedures for adjustment and correction. \textbf{Constellation Simulation (D.4, 1.13\%)} focuses on multi-satellite systems. It simulates constellation architectures and cooperative operational strategies. \textbf{Satellite System Simulation (D.5, 1.77\%)} models system-level capabilities and comprises two subtypes. \textit{Internal Subsystem Simulation (D.5.1, 1.29\%)} focuses on the design and operation of individual subsystems, such as communication and mission planning. \textit{External Interaction Simulation (D.5.2, 0.48\%)} examines satellite interactions with the external environment.
% , e.g., observation performance.

% \textbf{Satellite System Simulation (D.5, 1.77\%)} represents the second-largest proportion. It models specific system-level capabilities and includes two subtypes. \textit{Internal Subsystem Simulation (D.5.1, 1.29\%)} models the design and functioning of individual subsystems, such as communication, networking, and mission planning. \textit{External Interaction Simulation (5.2, 0.48\%)} focuses on how satellites interact with their external environment, including observation performance and detector pointing accuracy.

\finding{
% Representing 7.42\% of all projects, the \textit{Simulation} category includes five subcategories. Among them, satellite orbit simulation is the most prevalent (3.06\%), whereas satellite environment simulation is the least represented (0.32\%).
% Accounting for 7.42\% of all projects, the \textit{Simulation} category comprises five subcategories. Satellite orbit simulation is the most prevalent (3.06\%), whereas satellite environment simulation is the least represented (0.32\%).
\textit{Simulation} category accounts for 7.42\% of projects and includes five types. Orbit simulation is the most common (3.06\%), while environment simulation is the least (0.32\%).
}
\vspace{-3mm}

\subsection{Testing (E)}

The \textit{Testing} category includes projects that verify and evaluate the behavior, performance, accuracy, or robustness of satellite-related software and operational functions. Unlike \textit{Simulation}, which models satellite systems or environments under controlled conditions, \textit{Testing} focuses on assessing whether a software component, algorithm, or operational function satisfies expected requirements. This category accounts for 1.94\% of the surveyed projects and comprises four subcategories: system performance testing, algorithm effectiveness testing, instrument calibration testing, and software stability testing. \textbf{System Performance Testing (E.1, 0.81\%)} evaluates key performance metrics, such as networking efficiency and communication link quality. \textbf{Algorithm Effectiveness Testing (E.2, 0.81\%)} assesses the effectiveness of task-oriented algorithms, including image processing and orbit determination. E.1 and E.2 are the most prevalent. \textbf{Calibration Testing (E.3, 0.16\%)} examines the calibration accuracy of onboard instruments, while \textbf{Stability Testing (E.4, 0.16\%)} focuses on the stability of flight software.

\subsection{Tool (F)}

% The \textit{Tool} category mainly focuses on lightweight, functionality-specific, modular tools, with 451 out of 620 of all projects (72.74\%). This is the largest-percentage category in all categories of our taxonomy. This category is divided into tools of 5 subcategories, including data utility (F.1), satellite spatial analysis (F.2), orbit computation (F.3), communication management (F.4), and visualization (F.5). We will introduce these subcategories.

The \textit{Tool} category comprises modular and function-oriented tools. It accounts for 72.74\% of the projects. As the largest category in the taxonomy, it includes five subcategories.

% : data utility, satellite spatial analysis tool, orbit computation tool, communication management tool, and visualization. 

% We introduce these subcategories as follows.

\subsubsection{Data Utility (F.1, 53.71\%)} 

It covers the core functional pipeline of satellite data tools, spanning data collection to data processing. \textit{Data Collection (F.1.1, 4.68\%)} obtains satellite data through two strategies: \textit{Data Retrieval (F.1.1.1, 2.42\%)}, which accesses existing data via requirement-driven queries, metadata filtering, or catalog services; and \textit{Data Download (F.1.1.2, 2.26\%)}, which downloads raw or pre-processed satellite data from repositories or ground stations for further use. \textit{Data Processing (F.1.2, 49.03\%)} converts satellite data into application-ready information and includes two types: 

% non-imagery processing (F.1.2.1) and imagery processing (F.1.2.2).
% which are described below.

% We introduce F.1.2.1 and F.1.2.2, respectively, as follows.

\textbf{Non-Imagery Processing (F.1.2.1, 7.26\%)} focuses on the analysis of non-imagery satellite data, such as telemetry, sensor measurements, or radio-frequency signals, to derive quantitative parameters and geophysical variables. This subcategory is classified into two types.
% based on scenario-specific processing and scenario-independent processing. 
% \textbf{Application Domains (F.1.2.1.1, 5.16\%)} classify related OSS projects according to their areas, as shown in Fig.~\ref{fig:F1211}. A total of 32 projects are identified, distributed across four types of non-imagery data processing: environmental monitoring, astrometry data analysis, geolocation analysis, and others. Among them, 18 projects focus on environmental monitoring. They integrate multi-source satellite, radar, and sensor data to support weather and climate analysis (e.g., cloudburst forecasting, precipitation estimation), air quality assessment (e.g., $CO_2$, $NO_2$), wildlife tracking (e.g., sea-turtle spatial behavior), and oceanographic applications including ocean color monitoring and cryosphere change detection. 5 projects concern astronomy data analysis. Their functions include satellite band classification and fusion, variable-star light-curve analysis, gamma-ray burst interpretation, and lifetime estimation based on orbital decay models. 7 projects address geolocation analysis. They provide data positioning, GPS coordinate transformation, and user/ship/animal tracking on geographic maps. The remaining projects support other domains, spanning satellite reconnaissance data interpretation for potential target detection, and commercial Earth-observation market analyses for economic trend assessment. 
\textit{Scenario-Specific Processing (F.1.2.1.1, 5.16\%)} groups OSS projects by application scenario, spanning four types.
% : environmental monitoring, astronomy data analysis, geolocation analysis, and others. 
(1) Environmental monitoring dominates, integrating multi-source satellite, radar, and sensor data for weather and climate analysis, air quality assessment, wildlife tracking, and oceanographic studies. (2) Astronomy-related projects support spectral analysis, light-curve modeling, gamma-ray burst interpretation, and orbital-decay–based lifetime estimation. (3) Geolocation analysis focuses on GPS coordinate transformation and entity tracking. (4) The remaining projects address satellite reconnaissance data interpretation and Earth-observation market analysis.
% In total, 32 projects are identified across four types of non-imagery data processing: environmental monitoring, astrometry data analysis, geolocation analysis, and others. Environmental monitoring dominates with 18 projects, integrating multi-source satellite, radar, and sensor data for weather and climate analysis (e.g., precipitation estimation), air quality assessment (e.g., $CO_2$), wildlife tracking (e.g., sea-turtle spatial behavior), and oceanographic studies (e.g., cryosphere change detection). Astronomy-related projects (5) support tasks such as spectral band classification and fusion, variable-star light-curve analysis, gamma-ray burst interpretation, and orbital-decay–based lifetime estimation. Geolocation analysis comprises 7 projects, focusing on GPS coordinate transformation and entity tracking (e.g., users, vessels, animals). The remaining projects address other domains, including satellite reconnaissance data interpretation for target detection and commercial Earth-observation market analysis for economic trend assessment.
\textit{Scenario-Independent Processing (F.1.2.1.2, 2.10\%)} characterizes general stages in satellite data processing workflows, mainly including preprocessing and analysis. Preprocessing involves tasks such as data normalization, data fusion, and containerized execution. Analysis centers on spatiotemporal analysis, feature extraction, classification, and related tasks.

\begin{figure*}[t]
	\centering
    \includegraphics[width=0.92\textwidth]{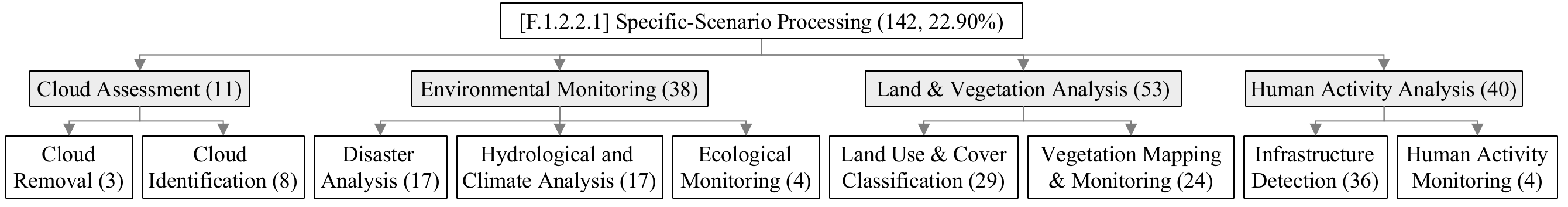}
    \vspace{-3mm}
    \caption{(RQ2) Scenario distribution in F.1.2.2.1. Numbers in parentheses denote the number of involved projects.}
    % \vspace{-1mm}
    \label{fig:F1221}
    \vspace{-3mm}
\end{figure*}

\textbf{Imagery Processing (F.1.2.2, 41.77\%)} targets the analysis of satellite imagery for diverse applications. It is also subdivided into two types: scenario-specific processing and scenario-independent processing. 
\textit{Scenario-Specific Processing (F.1.2.2.1, 22.90\%)} targets imagery analysis under different scenarios, involved in 142 projects (Figure~\ref{fig:F1221}). These projects fall into four primary scenarios. (1) Cloud assessment focuses on cloud removal and cloud identification, enabling cloud-free composite generation and low-contamination scene selection. (2) Environmental monitoring is based on imagery to cover disaster assessment (e.g., wildfire prediction and storm-surge forecasting), hydrological and climatic analysis (e.g., iceberg tracking and precipitation estimation), and ecological monitoring (e.g., invasive-species detection and methane ($CH_4$) emission analysis). (3) Land \& vegetation analysis involves land-use and land-cover interpretation, including surface-type classification, temporal change prediction, and vegetation monitoring such as tree-count estimation and health assessment. (4) Human activity analysis targets the extraction of human-related geographic features and socio-economic patterns, supporting automated detection of urban infrastructure (e.g., buildings and roads), long-term urban expansion monitoring, and estimation of indicators such as regional poverty levels.
% \textbf{General-Scenario Processing (F.1.2.2.2, 18.87\%)} addresses generic processing workflows for satellite imagery, covering 117 related projects. As shown in Fig.~\ref{fig:F1222}, these projects align with fundamental stages of the image-processing pipeline, including image preprocessing, image enhancement, image analysis, and image visualization. A portion of the tools also provides integrated, end-to-end functionality that spans multiple processing stages. Image preprocessing (25) involves fundamental operations such as coordinate reprojection, decoding, clipping, compression, denoising, radiometric calibration, image registration, and geometric transformation. These procedures standardize raw imagery and prepare it for downstream analytical tasks. Image enhancement (12/117) aims to improve visual quality or emphasize salient features, with common functions including super-resolution, dehazing, pansharpening, and synthetic image generation. Image analysis (48/117) represents the core analytical stage, encompassing semantic segmentation, classification, object detection, change detection, and regression to derive semantic and structural information from imagery. Image visualization (15/117) supports the rendering and interpretation of processed or raw imagery and facilitates the visual exploration of phenomena such as burned areas or high-resolution scenes. In addition, a set of comprehensive workflows (16/117) integrates multiple processing stages into unified pipelines, enabling flexible, modular, and end-to-end usage across diverse application contexts.
\textit{Scenario-Independent Processing (F.1.2.2.2, 18.87\%)} focuses on generic satellite imagery processing and includes 117 projects. These projects correspond to core stages: image preprocessing, enhancement, analysis, and visualization, with some tools offering integrated end-to-end workflows. Image preprocessing covers essential operations, e.g., coordinate reprojection, decoding, denoising, and geometric transformation to standardize raw imagery. Image enhancement improves visual quality or highlights salient features through techniques, e.g., super-resolution, dehazing, and synthetic image generation. Image analysis constitutes the central analytical stage, encompassing tasks, e.g., semantic segmentation, change detection, and regression for extracting semantic and structural information. Image visualization supports the rendering and exploration of raw or processed imagery, while comprehensive processing workflows integrate multiple stages into unified pipelines.

% \begin{figure*}[t]
% 	\centering
%     \includegraphics[width=0.8\textwidth]{pic/F1222.pdf}
%     \vspace{-3mm}
%     \caption{(RQ2) Stage distribution in F.1.2.2.2. Numbers in parentheses denote the number of involved projects.}
%     % \vspace{-1mm}
%     \label{fig:F1222}
%     \vspace{-3mm}
% \end{figure*}

\subsubsection{Satellite Spatial Analysis Tool (F.2, 4.35\%)}

This category addresses the extraction and interpretation of satellite-related spatial information and comprises three subtypes. \textbf{Satellite Coverage Analysis (F.2.1, 0.65\%)} examines the geographic areas observed or serviced by sensing instruments or communication footprints. \textbf{Satellite Geolocation Analysis (F.2.2, 3.06\%)} focuses on estimating and refining the spatial positions of satellite observations. \textbf{Satellite Scheduling (F.2.3, 0.65\%)} supports the prediction and planning of satellite overpasses for specified target regions.

% For \textit{Satellite Signal Analysis (F.2)}, it focuses on examining the properties and quality of signals emitted, received, or relayed by satellites. The goal is to ensure signal performance, detect interference, and support robust satellite communication and observation systems. 

% For \textit{Satellite Spatial Analysis (F.3)}, it aims to reveal the spatial information for satellites, such as \textit{Satellite Coverage Analysis (F.3.1)}, \textit{Satellite Geolocation Analysis (F.3.2)}, and \textit{Satellite Overpass Planning (F.3.3)}.
% % \textit{Satellite Coverage Analysis (F.3.1)} analyzes and models the geographical areas covered by satellite sensors or communication footprints, ensuring mission requirements for spatial reach and frequency of coverage are met.
% \textit{Satellite Geolocation Analysis (F.3.2)} determines the precise geographical positions of satellite observations or communication signals, enhancing accuracy for applications like positioning, navigation, and remote sensing.
% \textit{Satellite Overpass Planning (F.3.3)} predicts and schedules satellite overpasses over regions of interest to optimize data acquisition, minimize conflicts, and coordinate multi-satellite observations or operations.

\subsubsection{Orbit Computation Tool (F.3, 3.55\%)}

This subcategory focuses on satellite trajectory analysis and comprises two core operations. \textbf{Orbit Determination (F.3.1, 0.81\%)} estimates precise orbital elements from tracking data or optical observations. \textbf{Orbit Propagation (F.3.2, 2.74\%)} predicts future orbital states by integrating the equations of motion.

% such as atmospheric drag, gravitational variations, and solar radiation pressure.

% For \textit{Orbit Computation (F.4)}, it involves calculating and predicting satellite trajectories. This category is further divided into two primary subcategories \textit{Orbit Status \& Risk Analysis (F.4.1)} and \textit{Core Computation (F.4.2)}.
% \textit{Orbit Status \& Risk Analysis (F.4.1)} assesses the current state of satellite orbits and identifies potential risks, such as anomalies or collision threats, to support operational safety and mission continuity.
% \textit{Core Computation (F.4.2)} performs the fundamental mathematical and physical calculations required to model and predict satellite orbits accurately. According to task goals, core computation is divided into 4 subcategories: \textit{Orbit Determination (F.4.2.1)}, \textit{Orbit Prediction (F.4.2.2)}, \textit{Orbit Propagation (F.4.2.3)}, and \textit{Orbit Perturbation Analysis (F.4.2.4)}. 
% Orbit determination aims to derive precise orbital parameters from observational data (e.g., tracking measurements, optical data) to define a satellite’s current orbit.
% Orbit prediction aims to forecast future orbital paths, accounting for gravitational forces and perturbations, to enable mission planning and collision avoidance.
% Orbit propagation aims to calculate future orbital positions by numerically integrating orbital equations over time, considering forces like drag, gravity, and radiation pressure.
% Orbit perturbation analysis aims to analyze small external forces or events that alter a satellite’s orbit.

\subsubsection{Satellite Communication Management Tool (F.4, 2.90\%)}

This category includes tools for managing and optimizing communication subsystems and targets three management objectives. \textbf{Link Analysis (F.4.1, 0.65\%)} evaluates inter–satellite and satellite–ground links in terms of quality, throughput, and reliability. \textbf{Protocol Analysis (F.4.2, 0.65\%)} assesses communication protocols for correctness and compliance. \textbf{Signal Analysis (F.4.3, 1.61\%)} focuses on signal processing to analyze communication signal quality.

% For \textit{Satellite Communication Management (F.5)}, it addresses the design, evaluation, and optimization of satellite communication systems by providing the related tools. This category mainly includes two analysis goals \textit{Link Analysis (F.5.1)} and \textit{Protocol Analysis (F.5.2)}. 
% \textit{Link Analysis (F.5.1)} assesses the quality, capacity, and reliability of communication links between satellites and ground stations or between satellites.
% \textit{Protocol Analysis (F.5.2)} provides tools that evaluate communication protocols.

% For \textit{Satellite Network Management (F.6)}, it focuses on orchestrating and controlling complex satellite networks.

\subsubsection{Visualization (F.5, 8.23\%)}

This subcategory focuses on analytical results and data through intuitive visual representations. It comprises two aspects. \textbf{Satellite Trajectory Visualization (F.5.1, 5.65\%)} renders orbital paths and temporal motion of single or multiple satellites, facilitating understanding of spatial dynamics and mission behavior. \textbf{Satellite Data Visualization (F.5.2, 2.58\%)} visualizes analytical outputs, improving the data interpretability.

% For \textit{Visualization (F.7)}, it provides the transformation of analytical outputs into clear, insightful visual formats. This category is divided into 4 sub-tasks: \textit{Trajectory Visualization (F.7.1)}, \textit{Event Visualization (F.7.2)}, \textit{Satellite Data Visualization (F.7.3)}, and \textit{Other Visualization (F.4)}.
% \textit{Trajectory Visualization (F.7.1)} visually represents satellite orbits and movements over time, facilitating understanding of spatial dynamics and mission scenarios.
% \textit{Event Visualization (F.7.2)} depicts specific space events (e.g., XX) for rapid situational awareness and analysis. 
% \textit{Satellite Data Visualization (F.7.3)} presents satellite-derived data such as imagery, measurements, or thematic maps in intuitive visual formats for scientific and operational use.
% \textit{Other Visualization (F.4)} supports additional custom visualization needs, enabling tailored insights for diverse user requirements across satellite applications.

% \finding{Most projects (72.74\%) fall under the \textit{Auxiliary Tool} category, which spans a broad range of functions across 9 inner categories and 16 leaf categories, including \textit{Data Utility}, \textit{Satellite Spatial Analysis}, \textit{Orbit Computation},  \textit{Satellite Communication Management}, and \textit{Visualization}. \textit{Data Utility} constitutes the largest subcategory, representing 53.71\% of all OSS satellite projects. This subcategory primarily provides tools for data collection and data processing, encompassing both non-imagery and imagery workflows.
% }

\finding{
72.74\% of the projects belong to the \textit{Tool} category, the largest in the taxonomy, spanning 9 inner and 16 leaf categories. \textit{Data Utility} dominates this category (53.71\%), focusing on satellite data collection and processing for both non-imagery and imagery data.
}

\vspace{-3mm}

\section{RQ3: Current Development Practices}\label{sec:rq3}

% This section examines current development practices in OSS satellite projects. The analysis is based on projects with publicly available source code, with a focus on the programming languages and implementation methodologies employed. Among all categories in Fig.\ref{fig:taxonomy}, \textit{Conceptual Design (A)} and \textit{Data Acquisition (B)} do not involve source code–level context. Thus, we summarize the practice for the remaining categories.

This section examines development practices based on projects with available source code, with a focus on languages and implementations. The categories \textit{Conceptual Design} and \textit{Dataset} in Figure~\ref{fig:taxonomy} lack source code; thus, only the remaining categories are analyzed.

% We summarize current development practices in OSS satellite projects, including programming languages and implementation methodologies used in 620 projects with available source code. Among all categories, only \textit{Conceptual Design (A)} and \textit{Data Acquisition (B)} do not provide source code-level implementations. For the remaining categories, we present detailed practices.

% From the overall OSS projects with source code, there are 66 programming languages are used, suggesting widely interest from different aspects and XXX. Particularly, we further analyze the top languages, where the top 5 are Python, Jupyter Notebook, JavaScript, C++, and MATLAB.

\subsection{Programming Language Practices}

We identify 66 distinct languages, indicating substantial technological diversity and heterogeneous implementation preferences in satellite software. The six most prevalent languages are Python, Jupyter Notebook, JavaScript, C++, MATLAB, and HTML, accounting for 32.91\%, 28.06\%, 7.40\%, 4.40\%, 4.06\%, and 3.57\%, respectively. These results suggest that OSS satellite projects do not rely on a single dominant ecosystem, but instead combine scientific computing, system implementation, and Web-based development practices.

% span scripting, scientific computing, and systems-level development. Language distributions are further analyzed across categories.

% Focusing on the dominant portion of the distribution, as shown in Table~\ref{tab:OverallLanguages}, the top six languages are Python, Jupyter Notebook, JavaScript, C++, MATLAB, and HTML, indicating a blend of scripting, scientific computing, and systems-level development. We further analyze the language distribution for each category.

% \begin{figure}[t]
%     \centering
%     % \vspace{-3mm}
%     \begin{minipage}[t]{0.47\textwidth}
%         \centering
%         \includegraphics[width=\textwidth]{pic/OverallLanguages.pdf}
%         \caption{The language distribution of all OSS satellite projects with available source code.}
%         \label{fig:OverallLanguages}
%     \end{minipage}
%     \hfill
%    \begin{minipage}[t]{0.47\textwidth}
%         \centering
%         \includegraphics[width=\textwidth]{pic/OverallLanguages.pdf}
%         \caption{XXX.}
%         \label{fig:XXX}
%     \end{minipage}
%     % \vspace{-4mm}
% \end{figure}

\begin{figure*}[t]
	\centering
    \includegraphics[width=0.95\textwidth]{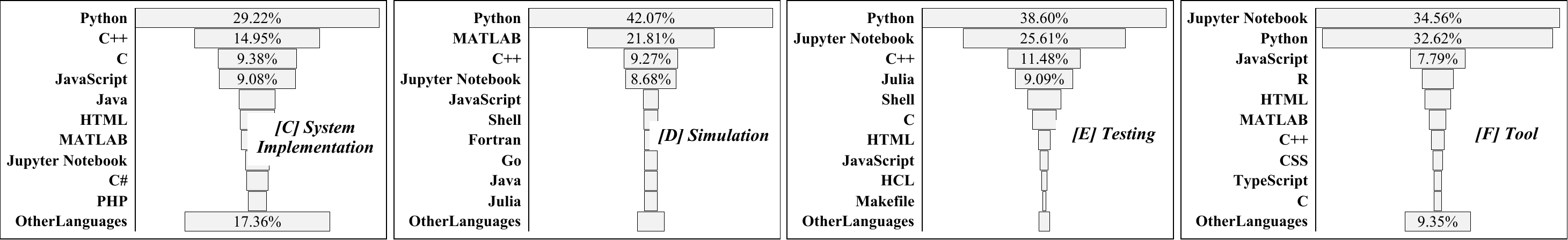}
    \vspace{-3mm}
    \caption{(RQ3) The language distribution (top 10) for each category of our taxonomy.}
    % \vspace{-1mm}
    \label{fig:LanguageDistributionEach}
    \vspace{-3mm}
\end{figure*}

% In \textit{System Implementation (C)} projects, Python (29.22\%) remains the dominant language, reflecting its suitability for satellite mission-logic development and rapid system prototyping. C++ (14.95\%) and C (9.38\%) together contribute over 24\%, showing the importance of low-level reliability and hardware-adjacent execution in onboard control, tracking, and subsystem management. JavaScript (9.08\%), Java (4.34\%), and HTML (4.17\%) primarily support visualization, network backends, and Web-based consoles, bridging ground systems with satellite runtime data. MATLAB (3.85\%) and Jupyter Notebook (2.86\%) indicate data-driven control models and system-level data processing, while C\# (2.56\%) and PHP (2.24\%) reflect legacy or platform-specific implementation requirements.

Language usage varies across categories, as shown in Figure~\ref{fig:LanguageDistributionEach}.
In \textit{System Implementation} (C), Python is the most common language (29.22\%), followed by C++ (14.95\%) and C (9.38\%). This indicates a combination of rapid satellite mission-logic development and low-level implementation needs in onboard control, tracking, and subsystem management.
JavaScript, Java, and HTML are also present, reflecting support for Web-based components.
%JavaScript, Java, and HTML are also present, reflecting support for service backends, interfaces, and Web-based control components.
In \textit{Simulation} (D), Python (42.07\%) and MATLAB (21.81\%) clearly dominate, with C++ used less frequently (9.27\%). This pattern shows that simulation projects primarily rely on high-level scientific computing environments, while a smaller portion requires performance-oriented implementations.
In \textit{Testing} (E), Python (38.60\%) and Jupyter Notebook (25.61\%) account for over 60\% of language usage. Compared with \textit{Simulation}, this category exhibits an emphasis on automation, experiment execution, and result inspection. C++ remains relevant (11.48\%), especially in testing performance-sensitive components, e.g., orbit computation and real-time communication systems.
In \textit{Tool} (F), Jupyter Notebook (34.56\%) and Python (32.62\%) are the two dominant languages. This is consistent with the exploratory and data-centric nature of these projects. JavaScript also appears frequently (7.79\%), mainly in visualization and interactive frontends.

\finding{
OSS satellite projects exhibit a heterogeneous, task-aligned programming landscape spanning 66 distinct languages. System-oriented projects use more low-level languages, simulation projects favor scientific computing environments, and testing and tool projects rely more heavily on interactive and scripting-based ecosystems.

% \textit{System Implementation} is dominated by Python, with C/C++ supporting performance-critical and hardware-proximal control. \textit{Simulation} and \textit{Testing} reinforce a scientific computing-oriented profile, relying on Python, MATLAB, Jupyter Notebook, and C++ for numerical modeling and experimental validation. \textit{Tool} shifts toward data-centric workflows, led by Jupyter Notebook and Python, with JavaScript and R supporting visualization and statistical analysis. Overall, mission-critical components prioritize determinism and execution stability, whereas analytical tools emphasize flexibility and interactivity. 
}
\vspace{-4mm}

% \finding{OSS satellite projects present a heterogeneous and task-aligned language landscape, covering 66 distinct languages. Meanwhile, mission-critical implementations emphasize determinism and execution stability, whereas analytical tools value flexibility and interactivity. Specifically, (1) \textit{System Implementation (C)} development predominantly uses Python, complemented by C/C++ for performance-sensitive and hardware-proximal control. \textit{Simulation (D)} and \textit{Testing (E)} further strengthen this scientific-computing profile, relying on Python, MATLAB, Jupyter Notebook, and C++ to support numerical modeling and experimental validation. In contrast, \textit{Auxiliary Tool (F)} transitions toward exploratory and data-centric workflows, led by Notebook-based processing and Python scripting, with JavaScript and R enhancing visualization and statistical analysis.

% OSS satellite projects exhibit a diverse language landscape. Mission-critical systems prioritize determinism and performance, while analytical tools prioritize flexibility, interactivity, and rapid iteration. Specifically, System-level development favors Python with substantial C/C++ usage for real-time and hardware-adjacent execution, whereas simulation and testing workloads rely heavily on scientific computing stacks dominated by Python, MATLAB, Notebook, and C++. In contrast, auxiliary tooling shifts toward exploratory, data-processing workflows led by Jupyter Notebook and Python, with JavaScript and R supporting visualization and analytical extensions.

\subsection{Implementation Practices}

We summarize implementation practices of the surveyed projects.

% across the categories in our taxonomy.

\subsubsection{System Implementation (C)} 
% This category is strongly satellite-dependent. It includes \textit{Mission-related Systems (C.1)} and \textit{Hardware-related Systems (C.2)}.
This category is highly satellite-specific, encompassing \textit{Mission-related Systems (C.1)} and \textit{Hardware-related Systems (C.2)}.

C.1 covers \textit{Satellite Control Systems (C.1.1)}, \textit{Communication and Network Systems (C.1.2)}, and \textit{Satellite-Ground Systems (C.1.3)}. 
C.1.1 includes \textit{Satellite Tracking (C.1.1.1)}, using SGP4 algorithms~\cite{p84} and TLE data for orbital state estimation, and \textit{Satellite Attitude Control (C.1.1.2)}, employing PD controllers or reinforcement learning for torque command generation. 
%$\bullet$ \textbf{C.1} comprises \textit{Satellite Control System (C.1.1)}, \textit{Communication and Network Systems (C.1.2)}, and \textit{Satellite-Ground Systems (C.1.3)}. Within C.1.1, two major branches are identified: \textit{Satellite Tracking (C.1.1.1)} and \textit{Satellite Attitude Control (C.1.1.2)}. For C.1.1.1, most implementations mainly estimate real-time orbital states from publicly available Two-Line Element (TLE) data, using the SGP4 propagation algorithm~\cite{p84}.
% \textbf{C.1.2} comprises \textit{Signal Processing (C.1.2.1)} for physical-layer decoding and antenna control; \textit{Communication Transmission (C.1.2.2)} for synchronization and Doppler compensation; and \textit{Network Routing (C.1.2.3)} for load-based congestion mitigation. 
%For C.1.1.2, projects focus on generating actuator commands to achieve desired control torques, mainly through fixed Proportional-Derivative (PD) controllers or reinforcement learning-based methods.
C.1.2 comprises \textit{Signal Processing (C.1.2.1)}, \textit{Communication Transmission (C.1.2.2)}, and  \textit{Network Routing (C.1.2.3)}. C.1.2.1 covers radio signal reception, decoding, and physical-layer/link-level data extraction, with some implementations further improving signal quality via antenna control based on fused sensor data and TLE orbital elements. C.1.2.2 mainly supports inter-device communication and time synchronization, while some projects enhance robustness through Doppler compensation and beam allocation optimization. C.1.2.3 addresses congestion by monitoring network load, classifying traffic by delay sensitivity, and applying differentiated rerouting.
% \textbf{C.1.3} comprises \textit{Satellite Mission Management (C.1.3.1)}, which supports operational scheduling, power management, and security monitoring, and \textit{Satellite-Generated Data Management (C.1.3.2)}, which covers SQL-based storage, telemetry visualization, and image analysis (e.g., wildfire and vegetation detection).
C.1.3 comprises \textit{Satellite Mission Management (C.1.3.1)} and \textit{Satellite-Generated Data Management (C.1.3.2)}. C.1.3.1 covers operational scheduling, power management, security monitoring, and real-time mission supervision. C.1.3.2 addresses the storage, analysis, and visualization of satellite-generated data, including SQL-based back-end support for data storage and querying, imagery analysis for vegetation, water-level, and wildfire monitoring, and visualization of spatial information, optical imagery, satellite events, and telemetry metrics.

\textit{\underline{Conclusion:} Mission-related systems combine orbit tracking and attitude control through both classical methods (e.g., SGP4 and PD control) and data-driven approaches such as reinforcement learning. They also support radio signal processing, robust communication, and adaptive routing in dynamic environments, while providing mission scheduling, power and system management, and end-to-end satellite data analysis, visualization, and storage for operational needs.
% Mission-related systems integrate orbit tracking and attitude control using both classical methods (e.g., SGP4 propagation and PD controllers) and data-driven approaches such as reinforcement learning. They further address radio signal processing, robust communication, and adaptive network routing under dynamic conditions. In addition, these systems support integrated mission scheduling, power and system management, as well as end-to-end analysis, visualization, and storage of satellite data for operational use.
% Mission-oriented satellite systems encompass orbit tracking and attitude control using both classical control techniques (e.g., SGP4-based propagation and PD controllers) and data-driven approaches such as reinforcement learning. They also address radio signal reception and decoding, robust communication, and network routing optimization under dynamic conditions. In addition, these systems support integrated mission scheduling, power management, system monitoring, and end-to-end processing, visualization, and storage of satellite data for operational and application-level use.
}

C.2 comprises \textit{Onboard Data Handling Systems (C.2.1)}, \textit{Integration Systems (C.2.2)}, and \textit{Control Systems (C.2.3)}. C.2.1 covers onboard data management implemented with custom PCBs, dedicated onboard computers, and prototyping platforms (Arduino). C.2.2 focuses on system integration, including CubeSat design, Eagle-based PCB layout, Linux-based operating platforms, and interface modules that connect sensors, communication units, and PCBs into integrated subsystems. C.2.3 is subdivided into \textit{Satellite Bus Control (C.2.3.1)} and \textit{Communication Signal Control (C.2.3.2)}. C.2.3.1 addresses tracking, attitude, power, thermal, and propulsion control using stepper motors, embedded platforms, accelerometer-based control with deep reinforcement learning, STM32-based power architectures, Arduino-based thermal regulation, and custom propulsion units. C.2.3.2 focuses on signal acquisition and processing using commercial receivers, DVB-S capture cards, IoT modules, Raspberry Pi, antenna/RF front-ends, and GNU Radio pipelines.

\textit{\underline{Conclusion:} Hardware-related satellite systems span PCB- and microcontroller-based onboard data management to CubeSat-level system integration, encompassing structural and embedded platform design. They further address satellite bus control, including power, attitude, thermal, and propulsion, and signal control through signal acquisition, RF front-end integration, and signal processing pipelines.}

\subsubsection{Simulation (D)}

This category comprises five subcategories: \textit{Satellite Environment Simulation (D.1)}, \textit{Satellite Attitude Simulation (D.2)}, \textit{Satellite Orbit Simulation (D.3)}, \textit{Constellation Simulation (D.4)}, and \textit{Satellite System Simulation (D.5)}.

% $\bullet$ D.1 focuses on thermal and electric field modeling. Thermal simulations combine solar radiation analysis with material properties (e.g., absorptivity and emissivity) to estimate temperature evolution. Electric field simulations solve the Laplace equation for charge distribution in vacuums to derive field configurations.
% \textit{\underline{Conclusion:} Satellite environment simulation implements thermal modeling based on solar radiation and material properties, as well as electric field modeling under vacuum conditions.}
 
D.1 focuses on thermal environment and electric field modeling. Thermal simulations estimate temperature evolution from solar radiation and key material properties, especially absorptivity and emissivity. Electric field simulations compute the vacuum field distribution by solving the Laplace equation for the charge distribution.
\textit{\underline{Conclusion:} Satellite environment simulation covers thermal modeling based on solar radiation and material properties, and vacuum electric field computation.}
% D.1 focuses on reproducing the satellites' physical environment. Thermal behavior modeling integrates solar radiation calculations with material parameters, primarily absorptivity and emissivity, to estimate temperature evolution. Electric field modeling solves the Laplace equation governing charge distribution in a vacuum to obtain the surrounding field configuration. 
% \textit{\underline{Conclusion:} Satellite environment simulation implement thermal modeling based on solar radiation and material properties, and vacuum electric field computation.}

% $\bullet$ D.2 focuses on attitude propagation and control. Propagation computes orientation via rigid-body dynamics, utilizing quaternion representations and variational integrators for numerical stability. Control generates torques to maintain orientations through closed-loop quaternion feedback, evaluating errors and applying corrective actions to ensure system stability.
D.2 focuses on attitude propagation and control. Attitude propagation models time-varying satellite orientation from rigid-body dynamics, using quaternion representations and variational integrators for numerical integration. Attitude control computes corrective torques through closed-loop quaternion feedback to achieve desired orientations with stability and accuracy.
\textit{\underline{Conclusion:} Satellite attitude simulation combines rigid-body attitude propagation with quaternion-based integration and closed-loop feedback control to predict orientation and generate corrective torques.}

D.3 addresses orbit propagation and control. Orbit propagation predicts satellite position and velocity using SGP4 and Keplerian dynamics, with the latter providing analytical motion under idealized assumptions. Orbit control models corrective maneuvers through thruster-generated forces that adjust orbital elements.
\textit{\underline{Conclusion:} Satellite orbit simulation combines orbit propagation and control, using SGP4 and Keplerian dynamics to predict motion and thruster-based inputs to regulate orbital elements.}

D.4 addresses constellation modeling and collaborative operations. Constellation modeling captures inter-satellite communication and system dynamics using TLE-based orbital configurations to evaluate architectures such as Walker, Street-of-Coverage, and Flower constellations. Collaborative operations support coordinated behaviors, including trajectory planning, formation control, and distributed approaches, e.g., federated learning in satellite swarms.
\textit{\underline{Conclusion:} Constellation simulation combines structural and dynamical modeling via inter-satellite communication and TLE-based orbital configurations, while supporting coordinated control, deployment planning, and distributed learning across satellite swarms.}
D.5 comprises \textit{Internal Subsystem Simulation (D.5.1)} and \textit{External Interaction Simulation (D.5.2)}. \textit{D.5.1} models satellite communication, networking, and mission planning. Communication simulation covers multibeam behavior, protocol-level operations such as Simple Serial Protocol and AX.25, and link-budget estimation. Networking simulation captures hybrid architectures, LEO/MEO topologies, and congestion under traffic bottlenecks. Mission planning simulation addresses high-level planning and scheduling within satellite systems. \textit{D.5.2} focuses on external interactions, particularly Earth observation, by modeling observation conditions and detector pointing to support sensing and imaging tasks.
\textit{\underline{Conclusion:} Satellite system simulation covers internal communication, networking, and mission planning, as well as external interactions, particularly Earth observation for sensing and imaging.}

\subsubsection{Testing (E)}
This category includes \textit{System Performance Testing (E.1)}, \textit{Algorithm Effectiveness Testing (E.2)}, \textit{Calibration Testing (E.3)}, and \textit{Stability Testing (E.4)}. 

% E.1 examines both network performance and communication link performance. Network performance testing evaluates Starlink’s access network across multiple protocol layers and diverse geographical settings, and compares the Starlink LEO network with terrestrial cellular systems. Communication link performance testing focuses on satellite downlink systems, analyzing how parameters such as antenna diameter, carrier frequency, and satellite altitude affect the bit error rate under various channel impairments.
% \textit{\underline{Conclusion:} Satellite system performance is evaluated through testing of network and communication links, assessing multi-layer protocol behavior, geographic variability, and the impact of key physical parameters on link reliability under diverse channel conditions.}

%For E.1, implementations evaluate network-level performance and communication link performance. Network performance testing examines the Starlink access network across multiple protocol layers and diverse geographical environments, with comparative analyses against terrestrial networks. Communication link performance testing focuses on satellite downlink, studying the impact of physical parameters, e.g., antenna aperture, carrier frequency, and satellite altitude, on bit error rate under various channel impairments.
E.1 evaluates network and communication link performance. Network testing compares Starlink access with terrestrial networks across protocol layers and geographic environments. Communication link testing examines satellite downlink performance, analyzing how parameters such as antenna aperture, carrier frequency, and satellite altitude affect bit error rate under different channel impairments.
\textit{\underline{Conclusion:} System performance testing assesses network and communication link performance by capturing multi-layer protocol behavior, geographic variability, and the effects of key physical parameters on link reliability under diverse channel conditions.}

E.2 evaluates mission-specific image-processing and estimation algorithms, including shoreline extraction, orbit determination, attitude estimation, and space situational awareness (SSA). Implementations assess orbit estimation under realistic sensor noise, validate deep-learning-based landmark detectors for attitude estimation, and compare SSA approaches under satellite-like computational and operational constraints.
\textit{\underline{Conclusion:} Algorithm effectiveness testing evaluates image-processing and estimation algorithms under realistic sensor noise and operational constraints, covering Earth observation, orbit and attitude determination, and space situational awareness.}

% E.3 focuses on the calibration of instrument functionalities. Typical practices include characterizing the polarization response of radio-telescope antenna beams, using satellite signals as reference sources to validate and refine system calibration parameters. \textit{\underline{Conclusion:} Instrument calibration is performed by characterizing antenna polarization responses and using satellite signals as reference sources to validate and refine calibration parameters.}

% For E.3, implementations of instrument functionality calibration focus on characterizing the polarization response of radio-telescope antenna beams. Satellite signals are employed as stable reference sources to adjust and refine system 
E.3 focuses on testing the polarization response of radio-telescope antenna beams. Satellite signals are employed as stable reference sources to adjust and refine system calibration parameters. \textit{\underline{Conclusion:} Calibration testing is achieved through antenna polarization characterization using satellite-based reference signals.}

% E.4 assesses the stability of flight software operating within a containerized environment. The evaluation issues command sequences, receives and stores telemetry data, and verifies the correctness and reliability of telemetry relay processes to ensure long-term software operation. 
% \textit{\underline{Conclusion:} Flight software stability is evaluated in a containerized environment by validating command execution and reliable telemetry handling for long-term operation.}

%For E.4, its implementations evaluate the stability of flight software in containerized environments by issuing command sequences, collecting telemetry data, and verifying the correctness of telemetry relay.
E.4 evaluates the stability of flight software in containerized environments by issuing commands, collecting and verifying telemetry data.
\textit{\underline{Conclusion:} Stability testing verifies flight software stability through reliable command execution and telemetry handling.}

\subsubsection{Tool (F)}

This category comprises \textit{Data Utility (F.1)}, \textit{Satellite Spatial Analysis Tool (F.2)}, \textit{Orbit Computation Tool (F.3)}, \textit{Satellite Communication Management Tool (F.4)}, and \textit{Visualization (F.5)}.

F.1 includes \textit{Data Collection (F.1.1)} and \textit{Data Processing (F.1.2)}. F.1.1 includes \textit{Data Retrieval (F.1.1.1)} and \textit{Data Download (F.1.1.2)}. 
%F.1.1.1 retrieves satellite imagery and geospatial data through three approaches: (i) publicly accessible APIs, including Google Maps, Bing Maps, MapTiler, Sentinel Hub, Leaf, satellites.io, LRGS, and space-track.org; (ii) data-driven retrieval using image similarity search based on Siamese neural networks or conventional image descriptors; and (iii) custom Web-scraping pipelines implemented with Selenium and BeautifulSoup. 
F.1.1.1 leverages open APIs, data-driven retrieval, and custom web-scraping. APIs include Google/Bing Maps, Sentinel Hub, and space-track.org. Data-driven methods deploy Siamese neural networks or image descriptors for similarity search. Web-scraping utilizes Selenium and BeautifulSoup. 
%F.1.1.2 enables multi-source and multi-format data download from remote-sensing satellites, NOAA archives, RSTN stations, and satellite constellations. It further provides access to LEO-based Sentinel imagery through platforms, e.g., SciHub, Copernicus Data Space Ecosystem, Google Earth Engine, Google Cloud Platform, and Amazon Earth on AWS via SpatioTemporal Asset Catalog APIs.
F.1.1.2 enables multi-source and multi-format data download from satellites, NOAA archives, RSTN stations, and satellite constellations. It further provides access to LEO-based Sentinel imagery through platforms, e.g., SciHub, Copernicus Data Space Ecosystem, Google Earth Engine, Google Cloud Platform, and Amazon Earth on AWS via SpatioTemporal Asset Catalog APIs.
F.1.2 is divided into \textit{Non-Imagery Processing (F.1.2.1)} and \textit{Imagery Processing (F.1.2.2)}. Their implementations are task-driven and scenario-specific; therefore, detailed practices are not elaborated here.
\textit{\underline{Conclusion:} Data utility supports multi-source retrieval and download of imagery and geospatial data from open APIs and remote-sensing platforms, together with task-oriented processing of both imagery and non-imagery data for geospatial and mission analysis.}

F.2 consists of \textit{Satellite Coverage Analysis (F.2.1)}, \textit{Satellite Geolocation Analysis (F.2.2)}, and \textit{Satellite Scheduling (F.2.3)}. F.2.1 determines accessible ground regions from the geometric intersection of satellite pointing vectors with the Earth’s surface. F.2.2 supports geolocation through pass-over monitoring, TLE/SGP4-based position determination, and optimization-based navigation adjustment. F.2.3 performs automated transmission scheduling under constraints such as access windows and link availability.
\textit{\underline{Conclusion:} Satellite spatial analysis tools support coverage assessment through pointing geometry, geolocation through orbit propagation and optimization, and automated transmission scheduling for accessible ground targets.}

F.3 consists of \textit{Orbit Determination (F.3.1)} and \textit{Orbit Propagation (F.3.2)}. F.3.1 estimates orbital elements from range, angular, and positional measurements. F.3.2 propagates orbital states forward or backward in time through numerical integration, using TLE-based elements, simplified perturbation models, SGP4, or ephemeris data parameterized by orbital elements.
\textit{\underline{Conclusion:} Orbit computation tools combine observation-based orbit determination with numerical orbit propagation to predict satellite states.}

F.4 comprises \textit{Link Analysis (F.4.1)}, \textit{Protocol Analysis (F.4.2)}, and \textit{Signal Analysis (F.4.3)}. F.4.1 performs link-budget analysis to assess signal attenuation and reception quality. F.4.2 supports protocol design and analysis, including packet parsing, integrity verification, and system state updating. F.4.3 covers signal processing tasks such as decryption, encoding, Doppler correction, and enhancement for downstream analysis.
\textit{\underline{Conclusion:} Satellite communication management tools combine link-budget evaluation, protocol design, and signal analysis and enhancement to support reliable transmission, data integrity, and signal quality.}

F.5 comprises \textit{Satellite Trajectory Visualization (F.5.1)} and \textit{Satellite Data Visualization (F.5.2)}. F.5.1 visualizes satellite trajectories in single- and multi-satellite settings, including Starlink and GNSS constellations. F.5.2 presents satellite-derived analysis results, including performance metrics, statistical outputs, decision-support results, metadata, and geolocation information.
\textit{\underline{Conclusion:} Visualization supports the representation of satellite trajectories and satellite-derived analytical results, enabling effective analysis and decision support in single- and multi-satellite contexts.}

\finding{
Implementation practices in OSS satellite projects range from hardware-related software design to mission-level software optimization, combining classical analytical models, numerical integration strategies, and data-driven methods. They increasingly reflect the synergy between physics-based modeling and machine-learning enhancement, with strong emphasis on hardware-grounded operational realism to satisfy real-world constraints and mission requirements.

% Implementation practices in OSS satellite projects span hardware-level design to mission-level optimization, encompassing multi-aspect system implementations, simulation, testing, and tool development. These practices integrate classical analytical models, numerical integration, and data-driven algorithms, reflecting a growing reliance on the synergy between physics-based modeling and machine-learning enhancement. Emphasis on operational realism ensures alignment with real-world constraints and mission requirements.
% The implementation practices of OSS satellite projects span the full spectrum from physical-level design to system-level modeling, from multiple-aspect simulations and testing to diverse tool development. These practices incorporate classical analytical models, numerical integration schemes, and emerging data-driven algorithms. This combination reflects a clear trend: satellite development increasingly depends on the synergy between physics-based modeling and machine-learning-driven enhancement. Moreover, these practices demonstrate a strong emphasis on operational realism, ensuring that development pipelines remain aligned with real-world constraints and mission requirements.

}

\vspace{-3mm}

% \subsubsection{Visualization (F.5)}
% For this subcategory, four related categories are involved: \textit{Satellite Trajectory Visualization (F.5.1)} and \textit{Satellite Data Visualization (F.5.2)}.

% For the subcategory \textit{Satellite Trajectory Visualization (F.5.1)}, most projects (20/32, XX\%) focus on visualizing the trajectory of an individual satellite (XX). The remaining projects (12/32, XX\%) provide visualization for multiple satellites of the same type, such as Starlink satellites, all satellites from a database, GNSS satellites, or entire constellations. \textbf{\ding{72} For satellite trajectory visualization: single-satellite trajectory visualization -> XX\% (XX) || multi-satellite trajectory visualization -> XX\% (XX).}

% For the subcategory \textit{Satellite Data Visualization (F.5.2)}, some projects (/, XX\%) visualize analysis results derived from satellite data, such as compression outcomes, plotted box figures, and data-driven decision-making outputs. Other projects (/, XX\%) focus on visualizing satellite-related properties and information, including active satellites in orbit, spectral response characteristics, general satellite information, and geolocation data. \textbf{\ding{72} For satellite data visualization: analysis result visualization -> XX\% (XX) || satellite information visualization -> XX\% (XX).}

\section{Discussion and Implications}\label{sec:implications}

We discuss our results and outline their actionable insights and research implications for satellite developers and satellite researchers.

\noindent $\bullet$ \textbf{Satellite Datasets.} Observation data (B.1) dominate OSS satellite dataset projects, enabling large-scale imagery datasets (Finding 3). This reflects the mission-critical role of Earth observation satellites and their focus on remote-sensing imagery collection. These datasets have supported extensive image analysis and learning-based research~\cite{hu2020image, de2021rainbench, yang2025fedclr, zhai2024seco}, providing clear benefits to the SE community by enabling reproducible research and shared benchmarks. 
However, dataset-oriented projects remain scarce overall, particularly those tailored to emerging onboard computing and in-orbit software execution scenarios~\cite{wen2025satelight, wang2021tiansuan}. As onboard image processing, autonomous execution, and in-orbit AI gain importance, this gap has become a critical bottleneck for both in-orbit intelligent application software development and empirical evaluation. \emph{Actionable insights and research implications follow.} For \underline{satellite developers}, these findings highlight the need for community-driven curation and standardization of datasets that reflect realistic onboard application workloads, thereby supporting the development, testing, and benchmarking of intelligent onboard software. For \underline{satellite researchers}, such datasets enable systematic studies of software architecture, performance optimization, and lifecycle management for in-orbit computing payloads. Overall, extending OSS satellite datasets beyond observation data toward application- and execution-oriented resources would provide a concrete basis for advancing satellite SE research and practice.

\noindent $\bullet$ \textbf{Software Configurations.}
Mission-related systems (C.1) dominate implementation projects, reflecting a shift toward data-centric, tightly integrated satellite–ground architectures (Finding 4). This reflects operational limitations, e.g., limited energy and intermittent communication. Increasing coordination and onboard data volumes make configuration management a key SE challenge. Traditional manual or source-code–centric configuration approaches~\cite{flederer2021configurable, mehta2020rex, sun2020testing, xu2013not} are increasingly impractical due to high maintenance overhead and limited adaptability. Given strict onboard constraints on computation, energy, and bandwidth, efficient configuration (e.g., acquisition rates, algorithm settings, and communication protocols) is essential, reflecting a shift toward configuration-driven and collaborative satellite software. \emph{Actionable insights and research implications follow.} For \underline{satellite developers}, these findings highlight the need for configurable architectures and runtime-adjustable parameter management to accommodate evolving mission requirements and dynamic satellite–ground coordination. For \underline{satellite researchers}, they point to opportunities in configuration modeling, validation, and testing for resource-constrained satellite–ground systems, supporting advances in configuration management frameworks and co-processing strategies for next-generation satellite software.

\noindent $\bullet$ \textbf{Thermal Environment Simulation.}
Within \textit{Simulation (D)}, orbit simulation (D.3) dominates, whereas environment simulation (D.1) is least represented (Finding 5). This reflects mission criticality. Orbit simulation dominates because reliable in-orbit operation is essential. Despite the importance of thermal effects for in-orbit reliability and performance, thermal environment simulation remains underexplored. Existing models mainly consider solar radiation and material properties~\cite{p112}, overlooking how thermal dynamics interact with onboard software execution under large temperature fluctuations caused by the lack of convection and repeated sunlight–eclipse transitions. As onboard computing intensifies, thermal behavior is increasingly influenced by software-controlled factors, e.g., task duration, workload intensity, and scheduling, which interact with quasi-periodic environmental cycles. \emph{Actionable insights and research implications follow.} For \underline{satellite developers}, incorporating software-aware thermal simulation into early design and validation can support thermal-conscious scheduling, workload shaping, and pre-deployment risk mitigation. For \underline{satellite researchers}, the quasi-periodic nature of space thermal cycles enables controlled models that incorporate execution behavior, offering reusable testbeds for thermal-aware software design, runtime adaptation, and performance–energy co-optimization.
Overall, advancing software-aware thermal simulation is essential for next-generation thermal-aware satellite software engineering.

\noindent $\bullet$ \textbf{Onboard Data Utilities.}
Within the \textit{Tools (F)} category, data utility (F.1) constitutes the largest subcategory (Finding 6), with most tools focused on data processing for ground-based use. This imbalance suggests limited support for in-orbit data processing despite improving onboard computing capabilities. In practice, satellite data often contain missing values, outliers, and duplicates, motivating AI-assisted quality assessment for in-orbit parsing, validation, and cleaning to reduce downlink overhead. \emph{Actionable insights and research implications follow.} For \underline{satellite developers}, migrating ground-side lightweight preprocessing functions (e.g., change detection, denoising, calibration, and image clipping) to onboard execution can reduce redundant data downlink and better mitigate resource constraints. For \underline{satellite researchers}, this opens opportunities to study in-orbit data quality modeling and data processing partitioning, supporting a shift from data–forwarding–centric to onboard-enhanced, quality-aware satellite software.

\noindent $\bullet$ \textbf{Open-Source Satellite Platforms.}
Figure~\ref{fig:taxonomy} shows that most OSS satellite projects target isolated components, such as data utilities, system modules, or standalone simulation environments. This suggests that the open-source satellite software ecosystem remains immature, with development largely concentrated on specialized tools and only partially open components. In addition, end-to-end satellite platforms are largely absent. This gap underscores the need for open-source whole-satellite platforms. Such platforms would support end-to-end experimentation, system-level evaluation, and cross-layer analysis of computation, communication, control, and environmental interactions. \emph{Actionable insights and research implications follow.} For \underline{satellite developers}, such platforms can be built by integrating best-practice open-source subsystems through modular architectures and well-defined interfaces rather than from scratch, and providing key infrastructure and realistic execution. For \underline{satellite researchers}, they enable reproducible system-level studies bridging simulation-centric evaluation and real in-orbit deployment. More broadly, they represent an important SE research direction, supporting rapid prototyping, comparative evaluation, and collaborative satellite software development.

\section{Threats to Validity}\label{sec:discussion}

\textbf{Researcher Subjectivity.}
This study relies on manual analysis to construct a taxonomy of satellite software goals, which may introduce researcher subjectivity and potentially affect validity. To mitigate it, two authors independently conduct the classification. Disagreements are resolved through discussion, with arbitration by a satellite-domain expert when necessary. Inter-rater reliability is assessed using Cohen's Kappa, yielding a score of 0.857, which indicates a high level of agreement and supports a reliable labeling process. To enhance transparency and reproducibility, all project classifications are publicly released in a repository~\cite{ourdata1, ourdata2}.

\noindent \textbf{Taxonomy Stabilization.}
When classifying the remaining 30\% of satellite projects in RQ2, four additional categories are identified and incorporated into the final taxonomy, which may raise concerns regarding completeness. However, the number of newly introduced categories is smaller than that reported in prior studies~\cite{chen2020comprehensive, WenServerless21}. Moreover, each new category contains no more than three projects, indicating limited expansion. Iterations mainly introduce minor category adjustments rather than new high-level categories. This suggests that our taxonomy has reached relatively stabilization. The final taxonomy is further reviewed by authors with satellite-domain expertise.

\section{Conclusion}\label{sec:conclusion}
This paper presents the first empirical study of open-source satellite software. Based on 22,286 satellite-related GitHub repositories, we examine popularity trends, functional objectives, and development practices. The results show sustained growth in both satellite software and active developers, underscoring the rising importance of open-source ecosystems in this domain. From a random sample of 646 projects, we derived a taxonomy of 43 software goals spanning datasets, system implementation, simulation, testing, and tools. We also find technological heterogeneity, with 66 programming languages used in task-aligned ways and increasing integration of analytical and data-driven models. Overall, the findings provide actionable insights for satellite developers and researchers.

\section{Data Availability Statement}
We have released all the data and code used for this study in online repositories~\cite{ourdata1, ourdata2}.

\begin{acks}
This work was partially funded by the National Natural Science Foundation of China under Grant No. 62425203, China Postdoctoral Science Foundation under Grant No. 2025M771560, and Fundamental Research Funds for the Central Universities under Grant No. 2024ZCJH11.
\end{acks}
\balance
\bibliographystyle{ACM-Reference-Format}
\bibliography{sample-base}

@misc{ourdata1,
  title = {SatelliteSoftwareStudy},
  howpublished = {\url{https://github.com/WenJinfeng/SatelliteSoftware}},
  note = {},
  year = 2026
}

@misc{ourdata2,
  title = {SatelliteSoftwareStudy},
  author ={Anonymous},
  doi={10.5281/zenodo.19247647},
  note = {},
  year = 2026
}

@misc{IntroStarlink,
  title = {Starlink},
  howpublished = {\url{https://www.starlink.com}},
  note = {},
  year = 2026
}

@misc{IntroOneWeb,
  title = {OneWeb: Connect with ease},
  howpublished = {\url{https://oneweb.net/connect_with_ ease}},
  note = {},
  year = 2026
}

@misc{IntroKuiper,
  title = {Amazon to offer broadband access from orbit with 3,236-satellite 'Project Kuiper' constellation},
  howpublished = {\url{https://www.geekwire.com/2019/amazonproject-kuiper-broadband-satellite/}},
  note = {},
  year = 2026
}

@misc{failurerate,
  title = {Small-Satellite Mission Failure Rates},
  howpublished = {\url{https://ntrs.nasa.gov/citations/20190002705}},
  note = {},
  year = 2026
}

@misc{diversesoftware,
  title = {State-of-the-art of small spacecraft technology},
  howpublished = {\url{https://www.nasa.gov/smallsat-institute/sst-soa/ground-data-systems-and-mission-operations/}},
  note = {},
  year = 2026
}

@misc{starlinkdetail,
  title = {Starlink satellites: Facts, tracking and impact on astronomy},
  howpublished = {\url{https://www.space.com/spacex-starlink-satellites.html}},
  note = {},
  year = 2026
}

@misc{earthobservation,
  title = {UCS Satellite Database},
  howpublished = {\url{https://www.ucsusa.org/resources/satellite-database}},
  note = {},
  year = 2026
}

@misc{githubapi,
  title = {GitHub REST API documentation},
  howpublished = {\url{https://developer.github.com/v3/}},
  note = {},
  year = 2026
}

@misc{satelliterepo,
  title = {Satellite-related Repository},
  howpublished = {\url{https://github.com/search?q=satellite&type=repositories}},
  note = {},
  year = 2026
}

@misc{p36,
  title = {Satellite clone},
  howpublished = {\url{https://github.com/RedHatSatellite/satellite-clone}},
  note = {},
  year = 2026
}

@misc{p112,
  title = {CubeSat thermal simulator},
  howpublished = {\url{https://github.com/TyeD/CubeSatThermal_Simulator}},
  note = {},
  year = 2026
}

@misc{p118,
  title = {OrbitPy},
  howpublished = {\url{https://github.com/EarthObservationSimulator/orbitpy}},
  note = {},
  year = 2026
}

@misc{p301,
  title = {openSatCon},
  howpublished = {\url{https://github.com/oygx210/openSatCon}},
  note = {},
  year = 2026
}

@misc{p20,
  title = {r2cloud},
  howpublished = {\url{https://github.com/dernasherbrezon/r2cloud}},
  note = {},
  year = 2026
}

@misc{p84,
    title = {sunflower-server},
    howpublished = {\url{https://github.com/PI2-sunflower/sunflower_server}},
    note = {},
    year = 2026
}

@misc{p310,
  title = {constellation-simulator},
  howpublished = {\url{https://github.com/GuilhemHnr/constellation-simulator}},
  note = {},
  year = 2026
}

@misc{p420,
  title = {CubeSat project},
  howpublished = {\url{https://github.com/ClouD-161803/CubeSat_project}},
  note = {},
  year = 2026
}

@misc{p285,
  title = {Semantic-Segmentation-Forestry},
  howpublished = {\url{https://github.com/Rohit8y/Semantic-Segmentation-Forestry}},
  note = {},
  year = 2026
}

@article{cohen1960coefficient,
  title   = {A coefficient of agreement for nominal scales},
  author  = {Jacob Cohen},
  journal = {Educational and psychological measurement},
  volume  = {20},
  number  = {1},
  pages   = {37--46},
  year    = {1960},
  DOI={10.1177/001316446002000104},
  publisher={Sage Publications Sage CA: Thousand Oaks, CA}
}

@article{landis1977measurement,
  title   = {The measurement of observer agreement for categorical data},
  author  = {J Richard Landis and Gary G Koch},
  journal = {Biometrics},
  number = {1},
  volume = {33},
  pages   = {159--174},
  year    = {1977},
  DOI={10.2307/2529310},
  publisher={Wiley, International Biometric Society}
}

@inproceedings{icseAghajaniNVLMBL19,
  author    = {Emad Aghajani and
               Csaba Nagy and
               Olga Lucero Vega{-}M{\'{a}}rquez and
               Mario Linares{-}V{\'{a}}squez and
               Laura Moreno and
               Gabriele Bavota and
               Michele Lanza},
  title     = {Software documentation issues unveiled},
  booktitle = {Proceedings of the International Conference on Software Engineering},
  pages     = {1199--1210},
  year      = {2019},
  doi={10.1109/ICSE.2019.00122}
  }

@inproceedings{CHENDLDEPLOY2,
  author    = {Zhenpeng Chen and
               Huihan Yao and
               Yiling Lou and
               Yanbin Cao and
               Yuanqiang Liu and
               Haoyu Wang and
               Xuanzhe Liu},
  title     = {An empirical study on deployment faults of deep learning based mobile applications},
  booktitle = {Proceedings of the International Conference on Software Engineering},
  pages = {674--685},
  year      = {2021},
  doi={10.1109/ICSE43902.2021.00068}
}

@inproceedings{isstaZhangCCXZ18,
  author    = {Yuhao Zhang and
               Yifan Chen and
               Shing{-}Chi Cheung and
               Yingfei Xiong and
               Lu Zhang},
  title     = {An empirical study on TensorFlow program bugs},
  booktitle = {Proceedings of the ACM SIGSOFT International Symposium on Software Testing and Analysis},
  pages     = {129--140},
  year      = {2018},
  doi = {10.1145/3213846.3213866}
}

@inproceedings{WenServerless21,
  author    = {Jinfeng Wen and
               Zhenpeng Chen and
               Yi Liu and
               Yiling Lou and
               Yun Ma and
               Gang Huang and
               Xin Jin and
               Xuanzhe Liu},
  title     = {An empirical study on challenges of application development in serverless computing},
  booktitle = {Proceedings of the ACM Joint Meeting on European Software Engineering Conference and Symposium on the Foundations of Software Engineering},
  pages = {416--428},
  year      = {2021},
  doi = {10.1145/3468264.3468558}
}

@inproceedings{bagherzadeh2019going,
  title={Going big: A large-scale study on what big data developers ask},
  author={Bagherzadeh, Mehdi and Khatchadourian, Raffi},
  booktitle={Proceedings of the ACM Joint Meeting on European Software Engineering Conference and Symposium on the Foundations of Software Engineering},
  pages={432--442},
  year={2019},
  doi = {10.1145/3338906.3338939}
}

@article{aghili2023studying,
  title={Studying the characteristics of AIOps projects on GitHub},
  author={Aghili, Roozbeh and Li, Heng and Khomh, Foutse},
  journal={Empirical Software Engineering},
  volume={28},
  number={6},
  pages={143},
  year={2023},
  DOI={10.1007/s10664-023-10382-z},
  publisher={Springer}
}

@article{ramasamy2023workflow,
  title={Workflow analysis of data science code in public GitHub repositories},
  author={Ramasamy, Dhivyabharathi and Sarasua, Cristina and Bacchelli, Alberto and Bernstein, Abraham},
  journal={Empirical Software Engineering},
  volume={28},
  number={1},
  pages={7},
  year={2023},
  doi={10.1007/s10664-022-10229-z},
  publisher={Springer}
}

@article{seaman1999qualitative,
  title={Qualitative methods in empirical studies of software engineering},
  author={Seaman, Carolyn B.},
  journal={IEEE Transactions on Software Engineering},
  volume={25},
  number={4},
  pages={557--572},
  year={1999},
  doi={10.1109/32.799955},
  publisher={IEEE}
}

@inproceedings{chen2020comprehensive,
  title={A comprehensive study on challenges in deploying deep learning based software},
  author={Chen, Zhenpeng and Cao, Yanbin and Liu, Yuanqiang and Wang, Haoyu and Xie, Tao and Liu, Xuanzhe},
  booktitle={Proceedings of the ACM Joint Meeting on European Software Engineering Conference and Symposium on the Foundations of Software Engineering},
  pages={750--762},
  doi = {10.1145/3368089.3409759},
  year={2020}
}

@inproceedings{lou2020understanding,
  title={Understanding build issue resolution in practice: symptoms and fix patterns},
  author={Lou, Yiling and Chen, Zhenpeng and Cao, Yanbin and Hao, Dan and Zhang, Lu},
  booktitle={Proceedings of the ACM Joint Meeting on European Software Engineering Conference and Symposium on the Foundations of Software Engineering},
  pages={617--628},
  year={2020},
  doi = {10.1145/3368089.3409760}
}

@inproceedings{humbatova2020taxonomy,
  title={Taxonomy of real faults in deep learning systems},
  author={Humbatova, Nargiz and Jahangirova, Gunel and Bavota, Gabriele and Riccio, Vincenzo and Stocco, Andrea and Tonella, Paolo},
  booktitle={Proceedings of the ACM/IEEE International Conference on Software Engineering},
  pages={1110--1121},
  doi = {10.1145/3377811.3380395},
  year={2020}
}

@article{garcia2012survey,
  title={A survey of discretization techniques: Taxonomy and empirical analysis in supervised learning},
  author={Garcia, Salvador and Luengo, Julian and S{\'a}ez, Jos{\'e} Antonio and Lopez, Victoria and Herrera, Francisco},
  journal={IEEE Transactions on Knowledge and Data Engineering},
  volume={25},
  number={4},
  pages={734--750},
  year={2012},
  doi={10.1109/TKDE.2012.35},
  publisher={IEEE}
}

@inproceedings{wang2021tiansuan,
  title={Tiansuan constellation: An open research platform},
  author={Wang, Shangguang and Li, Qing and Xu, Mengwei and Ma, Xiao and Zhou, Ao and Sun, Qibo},
  booktitle={Proceedings of the IEEE International Conference on Edge Computing},
  pages={94--101},
  year={2021},
  doi={10.1109/EDGE53862.2021.00022},
  organization={IEEE}
}

@inproceedings{wang2022quantifying,
  title={Quantifying community evolution in developer social networks},
  author={Wang, Liang and Li, Ying and Zhang, Jierui and Tao, Xianping},
  booktitle={Proceedings of the ACM Joint European Software Engineering Conference and Symposium on the Foundations of Software Engineering},
  pages={157--169},
  year={2022},
  doi = {10.1145/3540250.3549106}
}

@article{tamburri2019exploring,
  title={Exploring community smells in open-source: An automated approach},
  author={Tamburri, Damian A and Palomba, Fabio and Kazman, Rick},
  journal={IEEE Transactions on Software Engineering},
  volume={47},
  number={3},
  pages={630--652},
  year={2019},
  doi={10.1109/TSE.2019.2901490},
  publisher={IEEE}
}

@inproceedings{xing2024deciphering,
  title={Deciphering the enigma of satellite computing with COTS devices: Measurement and analysis},
  author={Xing, Ruolin and Xu, Mengwei and Zhou, Ao and Li, Qing and Zhang, Yiran and Qian, Feng and Wang, Shangguang},
  booktitle={Proceedings of the Annual International Conference on Mobile Computing and Networking},
  pages={420--435},
  year={2024},
  doi = {10.1145/3636534.3649371}
}

@inproceedings{michel2022first,
  title={A first look at Starlink performance},
  author={Michel, Fran{\c{c}}ois and Trevisan, Martino and Giordano, Danilo and Bonaventure, Olivier},
  booktitle={Proceedings of the ACM Internet Measurement Conference},
  pages={130--136},
  year={2022},
  doi = {10.1145/3517745.3561416}
}

@inproceedings{robic2022vision,
  title={Vision-based rotational control of an agile observation satellite},
  author={Robic, Maxime and Fraisse, Renaud and Marchand, Eric and Chaumette, Fran{\c{c}}ois},
  booktitle={Proceedings of the IEEE/RSJ International Conference on Intelligent Robots and Systems},
  pages={2211--2218},
  year={2022},
  doi={10.1109/IROS47612.2022.9981398},
  organization={IEEE}
}

@inproceedings{zhai2024seco,
  title={Seco: Multi-satellite edge computing enabled wide-area and real-time earth observation missions},
  author={Zhai, Zhiwei and Zeng, Liekang and Ouyang, Tao and Yu, Shuai and Huang, Qianyi and Chen, Xu},
  booktitle={Proceedings of the IEEE INFOCOM 2024-IEEE Conference on Computer Communications},
  pages={2548--2557},
  year={2024},
  doi={10.1109/INFOCOM52122.2024.10621270},
  organization={IEEE}
}

@inproceedings{zhang2024resource,
  title={Resource-efficient in-orbit detection of earth objects},
  author={Zhang, Qiyang and Yuan, Xin and Xing, Ruolin and Zhang, Yiran and Zheng, Zimu and Ma, Xiao and Xu, Mengwei and Dustdar, Schahram and Wang, Shangguang},
  booktitle={Proceedings of the IEEE INFOCOM 2024-IEEE Conference on Computer Communications},
  pages={551--560},
  year={2024},
  doi={10.1109/INFOCOM52122.2024.10621328},
  organization={IEEE}
}

@inproceedings{denby2023kodan,
  title={Kodan: Addressing the computational bottleneck in space},
  author={Denby, Bradley and Chintalapudi, Krishna and Chandra, Ranveer and Lucia, Brandon and Noghabi, Shadi},
  booktitle={Proceedings of the ACM International Conference on Architectural Support for Programming Languages and Operating Systems, Volume 3},
  pages={392--403},
  year={2023},
  doi = {10.1145/3582016.3582043}
}

@inproceedings{liu2024orbit,
  title={In-orbit processing or not? Sunlight-aware task scheduling for energy-efficient space edge computing networks},
  author={Liu, Weisen and Lai, Zeqi and Wu, Qian and Li, Hewu and Zhang, Qi and Li, Zonglun and Li, Yuanjie and Liu, Jun},
  booktitle={Proceedings of the IEEE INFOCOM 2024-IEEE Conference on Computer Communications},
  pages={881--890},
  year={2024},
  doi={10.1109/INFOCOM52122.2024.10621268},
  organization={IEEE}
}

@inproceedings{ma2023network,
  title={Network characteristics of Leo satellite constellations: A Starlink-based measurement from end users},
  author={Ma, Sami and Chou, Yi Ching and Zhao, Haoyuan and Chen, Long and Ma, Xiaoqiang and Liu, Jiangchuan},
  booktitle={Proceedings of the IEEE INFOCOM 2023-IEEE Conference on Computer Communications},
  pages={1--10},
  year={2023},
  doi={10.1109/INFOCOM53939.2023.10228912},
  organization={IEEE}
}

@inproceedings{pan2023pmsat,
  title={Pmsat: Optimizing passive metasurface for low earth orbit satellite communication},
  author={Pan, Hao and Qiu, Lili and Ouyang, Bei and Zheng, Shicheng and Zhang, Yongzhao and Chen, Yi-Chao and Xue, Guangtao},
  booktitle={Proceedings of the Annual International Conference on Mobile Computing and Networking},
  pages={1--15},
  year={2023},
  doi = {10.1145/3570361.3613257}
}

@inproceedings{singh2021community,
  title={A community-driven approach to democratize access to satellite ground stations},
  author={Singh, Vaibhav and Prabhakara, Akarsh and Zhang, Diana and Ya{\u{g}}an, Osman and Kumar, Swarun},
  booktitle={Proceedings of the Annual International Conference on Mobile Computing and Networking},
  pages={1--14},
  year={2021},
  doi = {10.1145/3447993.3448630}
}

@article{zhai2023fedleo,
  title={FedLEO: An offloading-assisted decentralized federated learning framework for low earth orbit satellite networks},
  author={Zhai, Zhiwei and Wu, Qiong and Yu, Shuai and Li, Rui and Zhang, Fei and Chen, Xu},
  journal={IEEE Transactions on Mobile Computing},
  volume={23},
  number={5},
  pages={5260--5279},
  year={2023},
  doi={10.1109/TMC.2023.3304988},
  publisher={IEEE}
}

@article{xie2024computation,
  title={Computation offloading and resource allocation in LEO satellite-terrestrial integrated networks with system state delay},
  author={Xie, Bo and Cui, Haixia and Ho, Ivan Wang-Hei and He, Yejun and Guizani, Mohsen},
  journal={IEEE Transactions on Mobile Computing},
  volume={24},
  number={3},
  pages={1372--1385},
  year={2024},
  doi = {10.1109/TMC.2024.3479243},
  publisher={IEEE}
}

@article{feng2024distributed,
  title={Distributed satellite-terrestrial cooperative routing strategy based on minimum hop-count analysis in mega LEO satellite constellation},
  author={Feng, Xin'ao and Sun, Yaohua and Peng, Mugen},
  journal={IEEE Transactions on Mobile Computing},
  volume={23},
  number={11},
  pages={10678--10693},
  year={2024},
  doi={10.1109/TMC.2024.3380891},
  publisher={IEEE}
}

@article{pan2022latency,
  title={Latency versus reliability in LEO mega-constellations: Terrestrial, aerial, or space relay?},
  author={Pan, Gaofeng and Ye, Jia and An, Jianping and Alouini, Mohamed-Slim},
  journal={IEEE Transactions on Mobile Computing},
  volume={22},
  number={9},
  pages={5330--5345},
  year={2022},
  doi={10.1109/TMC.2022.3168081},
  publisher={IEEE}
}

@inproceedings{mohan2024multifaceted,
  title={A multifaceted look at Starlink performance},
  author={Mohan, Nitinder and Ferguson, Andrew E and Cech, Hendrik and Bose, Rohan and Renatin, Prakita Rayyan and Marina, Mahesh K and Ott, J{\"o}rg},
  booktitle={Proceedings of the ACM Web Conference},
  pages={2723--2734},
  year={2024},
  doi = {10.1145/3589334.3645328}
}

@article{izhikevich2024democratizing,
  title={Democratizing LEO satellite network measurement},
  author={Izhikevich, Liz and Tran, Manda and Izhikevich, Katherine and Akiwate, Gautam and Durumeric, Zakir},
  journal={Proceedings of the ACM on Measurement and Analysis of Computing Systems},
  volume={8},
  number={1},
  pages={1--26},
  year={2024},
  doi = {10.1145/3639039},
  publisher={ACM New York, NY, USA}
}

@inproceedings{denby2020orbital,
  title={Orbital edge computing: Nanosatellite constellations as a new class of computer system},
  author={Denby, Bradley and Lucia, Brandon},
  booktitle={Proceedings of the International Conference on Architectural Support for Programming Languages and Operating Systems},
  pages={939--954},
  year={2020},
  doi = {10.1145/3373376.3378473}
}

@inproceedings{giuliari2021icarus,
  title={ICARUS: Attacking low earth orbit satellite networks},
  author={Giuliari, Giacomo and Ciussani, Tommaso and Perrig, Adrian and Singla, Ankit},
  booktitle={Proceedings of the USENIX Annual Technical Conference},
  pages={317--331},
  year={2021},
  url = {https://www.usenix.org/conference/atc21/presentation/giuliari}
}

@article{usman2020mitigating,
  title={Mitigating distributed denial of service attacks in satellite networks},
  author={Usman, Muhammad and Qaraqe, Marwa and Asghar, Muhammad Rizwan and Shafique Ansari, Imran},
  journal={Transactions on Emerging Telecommunications Technologies},
  volume={31},
  number={6},
  pages={e3936},
  year={2020},
  DOI={10.1002/ett.3936},
  publisher={Wiley Online Library}
}

@article{hu2020image,
  title={Image-based geo-localization using satellite imagery},
  author={Hu, Sixing and Lee, Gim Hee},
  journal={International Journal of Computer Vision},
  volume={128},
  number={5},
  pages={1205--1219},
  year={2020},
  doi={10.1007/s11263-019-01186-0},
  publisher={Springer}
}

@inproceedings{de2021rainbench,
  title={RainBench: Towards data-driven global precipitation forecasting from satellite imagery},
  author={de Witt, Christian Schroeder and Tong, Catherine and Zantedeschi, Valentina and De Martini, Daniele and Kalaitzis, Alfredo and Chantry, Matthew and Watson-Parris, Duncan and Bilinski, Piotr},
  booktitle={Proceedings of the AAAI Conference on Artificial Intelligence},
  volume={35},
  number={17},
  pages={14902--14910},
  year={2021},
  DOI={10.1609/aaai.v35i17.17749}
}

@inproceedings{legrand2022end,
  title={End-to-end neural estimation of spacecraft pose with intermediate detection of keypoints},
  author={Legrand, Antoine and Detry, Renaud and De Vleeschouwer, Christophe},
  booktitle={Proceedings of the European Conference on Computer Vision},
  pages={154--169},
  year={2022},
  DOI={10.1007/978-3-031-25056-9_11},
  organization={Springer}
}

@article{khalife2021first,
  title={The first carrier phase tracking and positioning results with Starlink LEO satellite signals},
  author={Khalife, Joe and Neinavaie, Mohammad and Kassas, Zaher M},
  journal={IEEE Transactions on Aerospace and Electronic Systems},
  volume={58},
  number={2},
  pages={1487--1491},
  year={2021},
  doi={10.1109/TAES.2021.3113880},
  publisher={IEEE}
}

@article{yang2025fedclr,
  title={FedCLR+: Tackling onboard label constraints for accurate federated satellite computing},
  author={Yang, Chen and Zhang, Qiyang and Sun, Qibo and Ouyang, Shufeng and Zhou, Ao and Wang, Shangguang and Xu, Mengwei},
  journal={IEEE Transactions on Services Computing},
  year={2025},
  volume={18},
  number={4},
  pages={2075-2088},
  doi={10.1109/TSC.2025.3583150},
  publisher={IEEE}
}

@inproceedings{yu2024comprehensive,
  title={A comprehensive analysis of security vulnerabilities and attacks in satellite modems},
  author={Yu, Lingjing and Hao, Jingli and Ma, Jun and Sun, Yong and Zhao, Yijun and Luo, Bo},
  booktitle={Proceedings of the ACM SIGSAC Conference on Computer and Communications Security},
  pages={3287--3301},
  year={2024},
  doi = {10.1145/3658644.3670390}
}

@inproceedings{wen2025satelight,
  title={SateLight: A satellite application update framework for satellite computing},
  author={Wen, Jinfeng and Zhao, Jianshu and Zhu, Zixi and Zhang, Xiaomin and Liang, Qi and Zhou, Ao and Wang, Shangguang},
  booktitle={Proceedings of the IEEE/ACM International Conference on Automated Software Engineering},
  year={2025},
  doi={10.1109/ASE63991.2025.00018}
}

@inproceedings{flederer2021configurable,
  title={A configurable framework for satellite software},
  author={Flederer, Frank and Montenegro, Sergio},
  booktitle={Proceedings of the IEEE International Conference on Software Engineering and Service Science},
  pages={28--31},
  year={2021},
  doi={10.1109/ICSESS52187.2021.9522190},
  organization={IEEE}
}

@inproceedings{shin2018test,
  title={Test case prioritization for acceptance testing of cyber physical systems: A multi-objective search-based approach},
  author={Shin, Seung Yeob and Nejati, Shiva and Sabetzadeh, Mehrdad and Briand, Lionel C and Zimmer, Frank},
  booktitle={Proceedings of the ACM SIGSOFT International Symposium on Software Testing and Analysis},
  pages={49--60},
  year={2018},
  doi = {10.1145/3213846.3213852}
}

@article{ollando2026test,
  title={Test schedule generation for acceptance testing of mission-critical satellite systems},
  author={Ollando, Rapha{\"e}l and Shin, Seung Yeob and Minardi, Mario and Sidiropoulos, Nikolas},
  journal={Empirical Software Engineering},
  volume={31},
  number={1},
  pages={1--35},
  year={2026},
  doi={10.1007/s10664-025-10737-8},
  publisher={Springer}
}

@inproceedings{esteve2012formal,
  title={Formal correctness, safety, dependability, and performance analysis of a satellite},
  author={Esteve, Marie-Aude and Katoen, Joost-Pieter and Nguyen, Viet Yen and Postma, Bart and Yushtein, Yuri},
  booktitle={Proceedings of the International Conference on Software Engineering},
  pages={1022--1031},
  year={2012},
  doi={10.1109/ICSE.2012.6227118},
  organization={IEEE}
}

@article{apvrille2004verifying,
  title={Verifying service continuity in a dynamic reconfiguration procedure: Application to a satellite system},
  author={Apvrille, Ludovic and de Saqui-Sannes, Pierre and S{\'e}nac, Patrick and Lohr, Christophe},
  journal={Automated Software Engineering},
  volume={11},
  number={2},
  pages={167--191},
  year={2004},
  DOI={10.1023/B:AUSE.0000017742.47984.6c},
  publisher={Springer}
}

@inproceedings{ozturk2023software,
  title={A software tool for planning LEO satellite operations},
  author={{\"O}zt{\"u}rk, Mustafa Yavuz and O{\u{g}}uzt{\"u}z{\"u}n, Halit},
  booktitle={Proceedings of the International Informatics and Software Engineering Conference},
  pages={1--6},
  year={2023},
  doi={10.1109/IISEC59749.2023.10391001},
  organization={IEEE}
}

@article{he2026enhancing,
  title={Enhancing the ability of LLMs for spaceborne equipment code generation via retrieval-augmented generation and contrastive learning},
  author={He, Rui and Zhang, Liang and Lyu, Liangqing and Xue, Changbin},
  journal={Automated Software Engineering},
  volume={33},
  number={1},
  pages={1--25},
  year={2026},
  doi={10.1007/s10515-025-00545-1},
  publisher={Springer}
}

@article{rico2016combined,
  title={A combined dependability and security approach for third party software in space systems},
  author={Rico, David Escorial and Hann, Mark},
  journal={arXiv preprint arXiv:1608.06133},
  year={2016},
  url={http://arxiv.org/abs/1608.06133}
}

@inproceedings{gios2024vision,
  title={A vision on a methodology for the application of an intrusion detection system for satellites},
  author={Gios, S{\'e}bastien and Bertrand Van Ouytsel, Charles-Henry and Carib{\'e}, Mark Diamantino and Legay, Axel},
  booktitle={Proceedings of the IEEE/ACM International Conference on Automated Software Engineering},
  pages={2205--2209},
  year={2024},
  doi = {10.1145/3691620.3695314}
}

@inproceedings{businge2019studying,
  title={Studying Android app popularity by cross-linking GitHub and Google Play Store},
  author={Businge, John and Openja, Moses and Kavaler, David and Bainomugisha, Engineer and Khomh, Foutse and Filkov, Vladimir},
  booktitle={Proceedings of the IEEE International Conference on Software Analysis, Evolution and Reengineering},
  pages={287--297},
  year={2019},
  doi={10.1109/SANER.2019.8667998},
  organization={IEEE}
}

@inproceedings{das2022empirical,
  title={An empirical study of blockchain repositories in GitHub},
  author={Das, Ajoy and Uddin, Gias and Ruhe, Guenther},
  booktitle={Proceedings of the International Conference on Evaluation and Assessment in Software Engineering},
  pages={211--220},
  year={2022},
  doi = {10.1145/3530019.3530041}
}

@inproceedings{trockman2019striking,
  title={Striking gold in software repositories? An econometric study of cryptocurrencies on GitHub},
  author={Trockman, Asher and Van Tonder, Rijnard and Vasilescu, Bogdan},
  booktitle={Proceedings of the IEEE/ACM International Conference on Mining Software Repositories},
  pages={181--185},
  year={2019},
  doi={10.1109/MSR.2019.00036},
  organization={IEEE}
}

@inproceedings{gonzalez2020state,
  title={The state of the ml-universe: 10 years of artificial intelligence \& machine learning software development on GitHub},
  author={Gonzalez, Danielle and Zimmermann, Thomas and Nagappan, Nachiappan},
  booktitle={Proceedings of the International Conference on Mining Software Repositories},
  pages={431--442},
  year={2020},
  doi = {10.1145/3379597.3387473}
}

@article{bao2019large,
  title={A large scale study of long-time contributor prediction for GitHub projects},
  author={Bao, Lingfeng and Xia, Xin and Lo, David and Murphy, Gail C},
  journal={IEEE Transactions on Software Engineering},
  volume={47},
  number={6},
  pages={1277--1298},
  year={2019},
  doi={10.1109/TSE.2019.2918536},
  publisher={IEEE}
}

@article{pickerill2020phantom,
  title={PHANTOM: Curating GitHub for engineered software projects using time-series clustering},
  author={Pickerill, Peter and Jungen, Heiko Joshua and Ochodek, Miros{\l}aw and Ma{\'c}kowiak, Micha{\l} and Staron, Miroslaw},
  journal={Empirical Software Engineering},
  volume={25},
  number={4},
  pages={2897--2929},
  year={2020},
  doi={10.1007/s10664-020-09825-8},
  publisher={Springer}
}

@inproceedings{mehta2020rex,
  title={Rex: Preventing bugs and misconfiguration in large services using correlated change analysis},
  author={Mehta, Sonu and Bhagwan, Ranjita and Kumar, Rahul and Bansal, Chetan and Maddila, Chandra and Ashok, Balasubramanyan and Asthana, Sumit and Bird, Christian and Kumar, Aditya},
  booktitle={Proceedings of the 17th USENIX Symposium on Networked Systems Design and Implementation},
  pages={435--448},
  year={2020},
  url = {https://www.usenix.org/conference/nsdi20/presentation/mehta}
}

@inproceedings{sun2020testing,
  title={Testing configuration changes in context to prevent production failures},
  author={Sun, Xudong and Cheng, Runxiang and Chen, Jianyan and Ang, Elaine and Legunsen, Owolabi and Xu, Tianyin},
  booktitle={Proceedings of the 14th USENIX Symposium on Operating Systems Design and Implementation},
  pages={735--751},
  year={2020},
  url = {https://www.usenix.org/conference/osdi20/presentation/sun}
}

@inproceedings{xu2013not,
  title={Do not blame users for misconfigurations},
  author={Xu, Tianyin and Zhang, Jiaqi and Huang, Peng and Zheng, Jing and Sheng, Tianwei and Yuan, Ding and Zhou, Yuanyuan and Pasupathy, Shankar},
  booktitle={Proceedings of the ACM Symposium on Operating Systems Principles},
  pages={244--259},
  year={2013},
  doi = {10.1145/2517349.2522727}
}

@article{calefato2022will,
  title={Will you come back to contribute? Investigating the inactivity of OSS core developers in GitHub},
  author={Calefato, Fabio and Gerosa, Marco Aurelio and Iaffaldano, Giuseppe and Lanubile, Filippo and Steinmacher, Igor},
  journal={Empirical Software Engineering},
  volume={27},
  number={3},
  pages={76},
  year={2022},
  doi={10.1007/s10664-021-10012-6},
  publisher={Springer}
}

@article{bock2023automatic,
  title={Automatic core-developer identification on GitHub: A validation study},
  author={Bock, Thomas and Alznauer, Nils and Joblin, Mitchell and Apel, Sven},
  journal={ACM Transactions on Software Engineering and Methodology},
  volume={32},
  number={6},
  pages={1--29},
  year={2023},
  doi = {10.1145/3593803},
  publisher={ACM New York, NY}
}


\end{document}